\documentclass[11pt]{article}
\usepackage{eurosym}
\usepackage{graphicx}
\usepackage[utf8]{inputenc}
\usepackage[T1]{fontenc}
\usepackage{indentfirst}
\usepackage{float}
\usepackage[margin=0.6in,nomarginpar]{geometry}
\usepackage[final]{hyperref}
\usepackage{amsmath}
\usepackage{hyperref}
\usepackage{cite}
\usepackage{physics}
\usepackage{subcaption}
\usepackage{caption}
\usepackage{amssymb}
\usepackage{multirow}
\usepackage[table]{xcolor}
\usepackage{orcidlink}
\usepackage{multirow,stackengine}
\usepackage{float}

\hypersetup{colorlinks=true, linkcolor=blue, citecolor=blue, urlcolor=blue}
\begin{document}

\title{Black Holes with Global Monopoles in 4D Noncommutative Einstein-Gauss–Bonnet Gravity}
\author{B. Hamil \thanks{%
hamilbilel@gmail.com } \\
Laboratoire de Physique Math\'{e}matique et Subatomique,\\
Facult\'{e} des Sciences Exactes, Universit\'{e} Constantine 1, Constantine,
Algeria. }
\date{\today }
\maketitle

\begin{abstract}
In this work, we construct an exact spherically symmetric black hole solution with a global monopole in the context of four-dimensional noncommutative Einstein-Gauss-Bonnet gravity. We modeled the spacetime noncommutativity via a Lorentzian-smeared mass distribution. Then we study the horizon structure and find that this black hole can have two configurations: one degenerate horizon or no horizon, depending on the black hole parameters. We also analyze thermodynamics and thermal stability by computing the Hawking temperature, entropy, and heat capacity. Our analysis reveals that the Hawking temperature and entropy acquire corrections from the noncommutative parameter $\Theta$, the energy scale of symmetry breaking $\eta$, and the Gauss-Bonnet coupling constant $\alpha$. The heat capacity exhibits divergences that signal second-order phase transitions. Thereafter, we study the black hole shadow employing the null geodesics and the Hamiltonian-Jacobi equation. Our results show that the shadow decreases with increasing $\Theta$ or $\alpha$ and increases with increasing $\eta$. Finally, we analyze quasinormal modes or scalar perturbations, we compute them via the 6th-order WKB method, and compare them to the shadow radius methods in the eikonal limit. 
\end{abstract}

\section{Introduction}
\label{sec1}
Modified gravity theories attracted significant interest as possible explanations for several unresolved astrophysical puzzles, such as the nature of dark matter, the formation of supermassive black holes (BHs),
galactic evolution, the dynamics of large-scale cosmic structures, and the accelerated expansion of the universe. Among these theories is Einstein-Gauss-Bonnet (EGB) gravity, which extends the standard gravitational theory by incorporating the Gauss-Bonnet (GB) invariant as an additional term in the gravitational Lagrangian. This gravitational theory can be viewed as an extension of General Relativity (GR) that includes quadratic curvature corrections, leading to significant implications for black hole physics, cosmology, and weak-field gravitational phenomena \cite{Fernandes}. The first such solution was introduced by Boulware and Deser in 1985 \cite{Tomozawa}, applicable to spacetimes with dimensions $D\geq 5$, since the GB term does not influence gravitational dynamics in four dimensions. However, later developments, notably by Tomozawa \cite{Cognola} and Cognola et al. \cite{Bousder}, demonstrated that through regularization techniques and dimensional reduction, the GB term can yield non-trivial effects even in 4-dimensional spacetimes. According to Lovelock's theorem, EGB gravity manifests itself only in dimensions $D\geq 5$, with the GB term lacking dynamic contribution in lower-dimensional spaces \cite{lin,Cognola}. This theoretical constraint prompted Glavan and Lin (2020) \cite{Gürses} to propose a coupling constant rescaling method to extract meaningful contributions to four-dimensional gravitational dynamics. Since then, considerable research has been dedicated to exploring the properties of the new 4D EGB theory \cite{Aoki,CLiu,Guo,Wei,Zubair,Malafarina,Mansoori,XHGe,Rayimbaev,Chakraborty,Odintsov,KYang,BAhmedov,RKumar,SGGhosh,Larranaga}

Global monopoles are topological defects formed during phase transitions in the early universe. Initially proposed by Barriola and Vilenkin \cite{ManuelBarriola}, these self-gravitating singularities emerge as static solutions in systems involving a triplet of real scalar fields, which transform under the fundamental representation of a global $O(3)$  symmetry group. This symmetry is spontaneously broken to $O(3)$ via a nonzero vacuum expectation value $\eta$. In the context of general relativity, Barriola and Vilenkin derived a metric describing a static black hole containing a global monopole, using a Schwarzschild background. The presence of the monopole modifies the black hole's topological structure compared to a standard Schwarzschild black hole, and its physical characteristics have been the subject of extensive study in recent years \cite{Lissa,Jiliang,Nikos,bilelhamil,Bronnikov,Valdir,KeJian,Fan}.

Noncommutative (NC) geometry arises naturally in the framework of open string theory. In particular, NC black holes play a significant role in the study of string theory and M-theory \cite{Mtheory}. Recent studies have demonstrated that the gravitational wave event GW150914 can be used to put constraints on quantum-scale fluctuations in NC spacetime \cite{spacetime}. This signal was recently detected by the LIGO and Virgo collaborations \cite{BPAbbot}. Moreover, it has been shown that the leading NC correction to the gravitational wave phase, generated by a binary system, appears at the second post-Newtonian order \cite{spacetime}.

Historically, the concept of NC spacetime was first introduced by Snyder \cite{Snyder} as a way to address divergences in relativistic quantum field theory. The central idea of NC frameworks is that spacetime coordinates, when treated as operators, no longer commute. In its simplest form, for a D-dimensional space, this noncommutativity can be written explicitly as:
\begin{equation}
\left[ x^a,x^b\right] =i\Theta ^{a b },
\end{equation}%
where $\Theta ^{a b }$ represents an antisymmetric matrix with dimensions of (length)$^{2}$. Several formulations of noncommutative field theory employ the Weyl-Wigner-Moyal *-product \cite{WeylH,EWigner}, but
these often fail to address significant issues such as the breaking of Lorentz invariance, the loss of unitarity, and ultraviolet divergences in quantum field theory. Recently, Smailagic and Spallucci \cite{Smailagic} proposed a novel approach called the coordinate coherent states framework, which offers a potential solution to these problems. In this model, objects such as a point mass $M$ are not treated as perfectly localized entities; instead, they are represented as spatially "smeared" distributions over a
region characterized by a length scale $\sqrt{\Theta }.$

In this study, we construct a 4D EGB black hole with global monopoles inspired by noncommutative geometry, where noncommutativity is modeled using a Lorentzian distribution in the mass. We explore not only the structure of the noncommutative solutions but also the thermodynamic stability of the system. Additionally, we provide a detailed analysis of the shadow radius, quasinormal modes, and greybody factors. 

This paper is organized as follows: In Section \ref{sec2}, we derive a black hole solution within the framework of 4D EGB gravity, incorporating a global monopole in a noncommutative spacetime, and study the structure of its horizons. In Section \ref{sec3}, we explore the thermodynamic properties of the black hole and discuss its stability. Section \ref{sec4} is dedicated to the investigation of the black hole shadow. Section \ref{sec5} focuses on the analysis of quasinormal modes (QNMs).  Finally, Section \ref{sec6} provides a summary of our results.

\section{Black hole solution}

\label{sec2}
In this section, we present the metric solution of the noncommutative 4D-EGB black hole with a global monopole. In this context, the action for the gravity model incorporating a global monopole is given by \cite{Cognola}:

\begin{equation}
S=\int d^{D}x\sqrt{-g}\left[ \frac{1}{16\pi }\left( R+\alpha%
\mathcal{G}\right) +\mathcal{L}^{\left( \mathrm{GM}\right) }\right] .
\end{equation}%
where $g$ is the metric determinant, $R$ is the Ricci scalar and $\alpha$ is the Gauss-Bonnet parameter of dimension [length]$^{2}$. Thethe GB quadratic curvature correction given by

\begin{equation}
\mathcal{L}^{\left( \mathrm{GM}\right)}=R^{2}-4 R_{\mu \nu }R^{\mu \nu }+R_{\mu \nu \sigma \delta }R^{\mu
\nu \sigma \delta },
\end{equation}

where $R_{\mu \nu }$, and $R_{\mu \nu \sigma \delta }$ are the Ricci tensor and the Riemann tensor, respectively. The global monopole Lagrangian is given by \cite{Fan}, 
\begin{equation}
\mathcal{L}^{\left( \mathrm{GM}\right) }=\frac{1}{2}\partial _{\mu }\chi
^{a}\partial ^{\mu }\chi ^{a}-\frac{\lambda }{4}\left( \chi ^{a}\chi
^{a}-\eta ^{2}\right) ^{2},\qquad a=1,2,...,D-1.
\end{equation}%
Here, $\lambda $ is the self-interaction constant and $\eta $ is the energy scale of symmetry breaking. The Einstein field equations are given by

\begin{equation}
G_{\mu \nu }+\alpha H_{\mu \nu }=8\pi \mathcal{T}_{\mu \nu },
\label{Einstein}
\end{equation}%
where $G_{\mu \nu }=R_{\mu \nu }-\frac{1}{2}g_{\mu \nu }R$ is the Einstein tensor, and $H_{\mu \nu }^{\left( \mathrm{GB}\right) }$ is the Lanczos tensor \cite{Lanczos} which is given by

\begin{equation}
H_{\mu \nu }=2\left( RR_{\mu \nu }-R_{\mu \sigma \delta \lambda }R_{\nu
}^{\delta \lambda \sigma }-2R_{\mu \sigma \nu \delta }R^{\sigma \delta
}-2R_{\mu \sigma }R_{\nu }^{\sigma }\right) -\frac{1}{2}\mathcal{L}^{\left( \mathrm{GM}\right)}%
g_{\mu \nu },
\end{equation}

and $\mathcal{T}_{\mu \nu }$ is the total energy-momentum tensor,
\begin{equation}
\mathcal{T}_{\mu \nu }=T_{\mu \nu }^{\left( \mathrm{NC}\right) }+T_{\mu \nu
}^{\left( \mathrm{GM}\right) }.
\end{equation}

Before calculating the metric of the $4D$ noncommutative EGB black hole with a global monopole, it is important to note that the regularization method proposed in Refs. \cite{Glavan,Tomozawa,Cognola} has been the subject of
debate \cite{Wen,Hennigar,Metin,Mahapatra}, and several questions have been raised regarding its validity. Other regularizations have also been proposed \cite{Hennigar,Kobayashi,Alessandro}, and among these, the regularization in \cite{Kobayashi,Alessandro} yields a well-defined scalar-tensor theory belonging to the Horndeski class. Moreover, the spherically symmetric $4D$ black hole solutions originally derived in Refs.\cite{Glavan,Tomozawa,Cognola} remain valid within these frameworks \cite{Hennigar,Kobayashi,Alessandro}. Therefore, it can be concluded that all of these regularization methods yield the same black hole solutions in the case of spherically symmetric $4D$ spacetimes. For simplicity and consistency, we will adopt the regularization method introduced in Ref. \cite{Glavan}.

The metric for a $D$-dimensional spherically symmetric black hole can be expressed in the form:
\begin{equation}
ds^{2}=-\mathcal{F}\left( r\right) dt^{2}+\frac{1}{\mathcal{F}\left(
r\right) }dr^{2}+r^{2}d\Omega _{D-2}^{2},
\end{equation}

where%
\begin{equation}
d\Omega _{D-2}^{2}=d\theta _{D-3}^{2}+\sin \theta _{D-3}^{2}\left( d\theta
_{D-4}^{2}+\sin \theta _{D-4}^{2}\left( ...+\sin \theta _{2}^{2}\left(
d\theta _{1}^{2}+\sin \theta _{1}^{2}d\varphi \right) \right) \right) .
\end{equation}%
In accordance with \cite{Nicollini2005,Farook,Arraut,Hamil,BHamil}, the noncommutative energy-momentum tensor $T_{\mu
\nu }^{\left( \mathrm{NC}\right) }$ is assumed to take the following form:%
\begin{equation}
T_{\mu }^{\nu \left( \mathrm{NC}\right) }=diagonal\left( -\rho \left(
r\right) ,p_{r},p_{\theta _{1}},...,p_{\theta _{D-3}},p_{\varphi }\right) ,
\end{equation}

where%
\begin{equation}
p_{r}=-\rho \left( r\right) ,
\end{equation}

and%
\begin{equation}
p_{\theta _{i}}=p_{\varphi }=-\rho \left( r\right) -\frac{r}{D-2}\frac{%
\partial }{\partial r}\rho \left( r\right) ,
\end{equation}%
while the energy-momentum tensor for the global monopole takes the form \cite{Fan}:%
\begin{equation}
T_{0}^{0\left( \mathrm{GM}\right) }=T_{r}^{r\left( \mathrm{GM}\right) }=%
\frac{\left( D-2\right) \eta ^{2}}{2r^{2}},T_{i}^{i\left( \mathrm{GM}\right)
}=\frac{\left( D-4\right) \eta ^{2}}{2r^{2}}\text{ \ where }i=2,3,...,D.
\end{equation}%
Rescaling the coupling constant as $\alpha \rightarrow \frac{\alpha }{D-4}
$, and taking the limit $D\rightarrow 4$, the ($r$,$r$)
component of Eq. (\ref{Einstein}) simplifies to%
\begin{equation}
2\left( r^{3}-2\alpha r\left( \mathcal{G}\left( r\right) -1\right) \right) 
\mathcal{G}\left( r\right) ^{\prime }+2\alpha \mathcal{G}\left( r\right)
^{2}+2r^{2}\mathcal{G}\left( r\right) -8\pi \xi \eta ^{2}r^{2}+8\pi \rho
\left( r\right) r^{4}=0,  \label{17}
\end{equation}%
where $\mathcal{G}\left( r\right) =1-\mathcal{F}\left( r\right) $. Here, we
adopt the mass distribution described by the following Lorentzian
distributions in 4D spacetime:%
\begin{equation}
\rho \left( r\right) =\frac{M\sqrt{\Theta }}{\pi ^{3/2}\left( r^{2}+\pi
\Theta \right) ^{2}}.
\end{equation}%
Using these energy density expressions and with an appropriate choice of the integration constant, Eq. (\ref{17}) admits the following solution

\begin{equation}
\mathcal{F}\left( r\right) =1+\frac{r^{2}}{2\alpha }\left( 1\pm \sqrt{\frac{%
32\alpha \eta ^{2}\pi }{r^{2}}+\frac{8M\alpha }{\pi r^{3}}\left( \pi -2\cot
^{-1}\left( \frac{r}{\sqrt{\pi \Theta }}\right) \right) -\frac{16\alpha 
\sqrt{\pi \Theta }M-\Theta \pi ^{2}r^{2}-\pi r^{4}}{\pi r^{2}\left(
r^{2}+\pi \Theta \right) }}\right) .
\end{equation}

To simplify, we expand the metric function to include first-order corrections from noncommutativity, resulting in:%
\begin{equation}
\mathcal{F}\left( r\right) \simeq 1+\frac{r^{2}}{2\alpha }\left( 1\pm \sqrt{%
1+\frac{4\alpha }{r^{2}}\left( \frac{2M}{r}+8\pi \eta ^{2}-\frac{8M}{\sqrt{%
\pi }r^{2}}\sqrt{\Theta }\right) }\right) .  \label{sol}
\end{equation}

This solution describes the noncommutative 4D EGB black hole, incorporating a global monopole. The solution is characterized by the GB coupling constant $\alpha $, the mass $M$, the energy scale of symmetry breaking $\eta $, and the noncommutative parameter $\Theta$. 

The solution reduces to the black hole case derived in \cite{Nikos} when $\Theta =0$. Additionally, if the GB coupling constant $\alpha =0$, the solution with positive sign leads to the Schwarzschild solution with an unphysical negative mass. In contrast, the solution with the negative sign correctly recovers the physically meaningful black hole solution. 

The horizon structure is determined by the roots of the equation $\mathcal{F}\left( r\right)=0$, which has a straightforward solution

\begin{equation}
r_{\pm }=\frac{1}{1-8\pi  \eta ^{2}}\left( M\pm \sqrt{M^{2}+8\pi \alpha
\xi \eta ^{2}-8M\sqrt{\frac{\Theta }{\pi }}\left( 1-8\pi \eta
^{2}\right) -\alpha }\right) .
\end{equation}

The largest root corresponds to the event horizon. Here, $r_{+}$ represents the event horizon, and $r_{-}$ denotes the Cauchy horizon. A simple examination of the zeros of the event horizon reveals the presence of a critical mass,

\begin{equation}
M_{c}=4\sqrt{\frac{\Theta }{\pi }}\left( 1-8\pi \xi \eta ^{2}\right) + 
\sqrt{\left( 1-8\pi \xi \eta ^{2}\right) \left( \alpha +\frac{16\Theta }{\pi 
}\left( 1-8\pi  \eta ^{2}\right) \right) }.
\end{equation}

\begin{figure}[H]
\begin{minipage}[t]{0.35\textwidth}
        \centering
        \includegraphics[width=\textwidth]{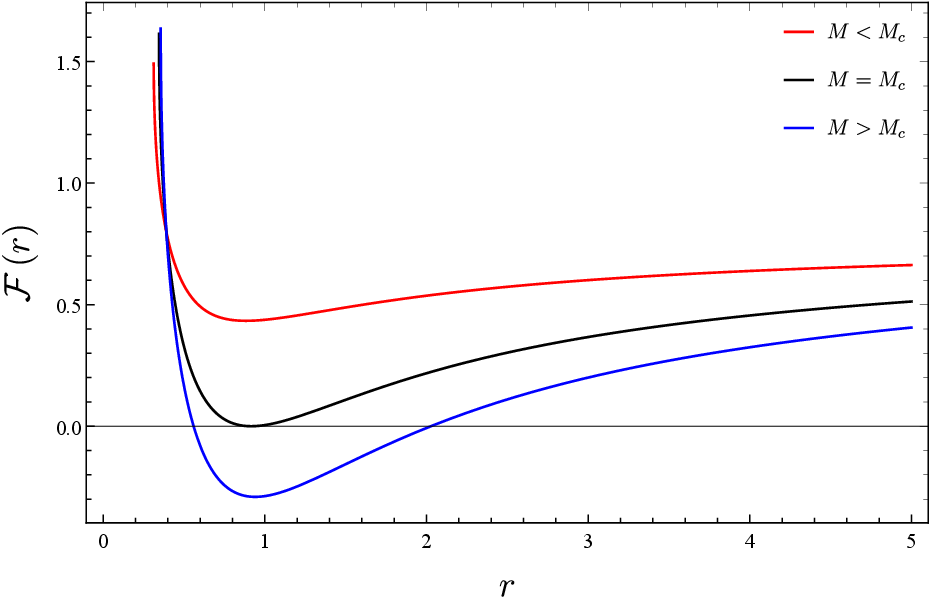}
       \subcaption{ $\alpha=0.095$, $\eta=0.095$, $\Theta=0.03$}\label{fig:met1}
   \end{minipage}%
\begin{minipage}[t]{0.350\textwidth}
        \centering
       \includegraphics[width=\textwidth]{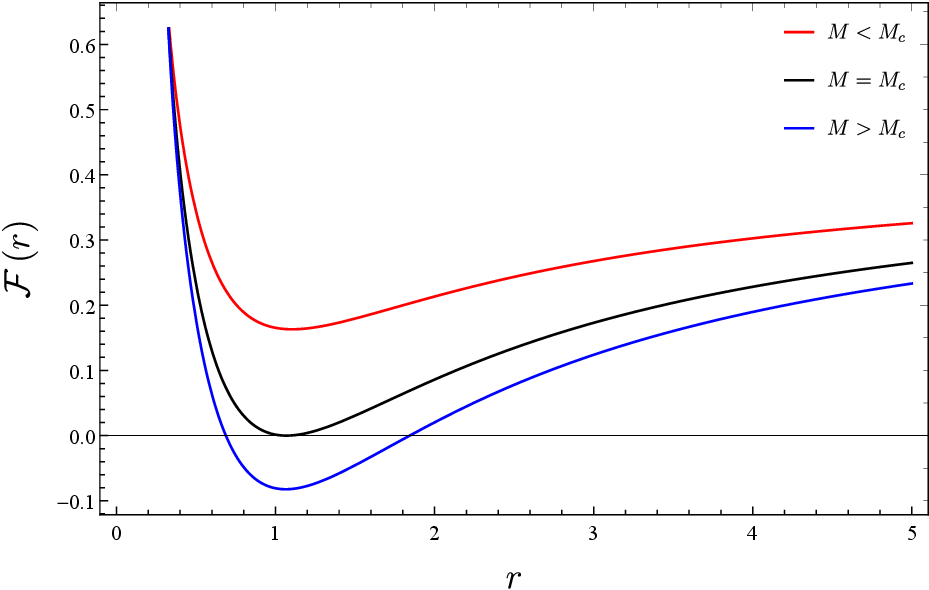}
        \subcaption{$\alpha=0.2$, $\eta=0.15$, $\Theta=0.02$}\label{fig:met2}
    \end{minipage}
\begin{minipage}[t]{0.35\textwidth}
     \centering
       \includegraphics[width=\textwidth]{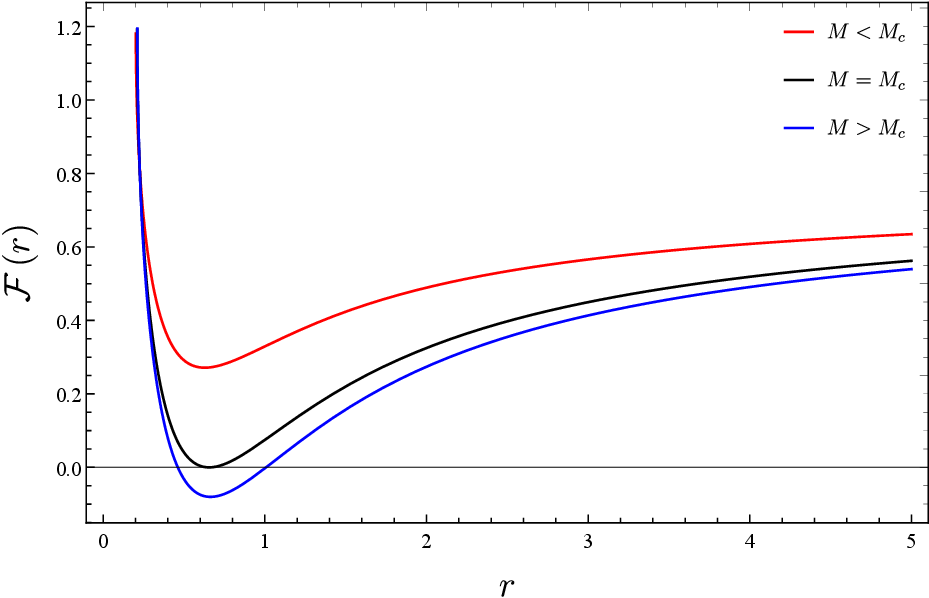}
        \subcaption{$\alpha=0.1$, $\eta=0.1$, $\Theta=0.01$}\label{fig:met3}
    \end{minipage}\hfill
\caption{Plot of the metric function $\mathcal{F}\left( r\right)$ vs. $r$}
\label{fig:met}
\end{figure}

When $M<M_{c}$, the equation $\left. \mathcal{F}\left( r\right)
\right\vert_{r=r_{+}}=0$ has no real roots; for $M_{c}=M$, it has a double root; and when $M_{c}<M$, the equation produces two different simple roots. Figure \ref{fig:met} illustrates the graph used to estimate the horizon. Contrary to a single event horizon, the figure explores three potential configurations for the black hole:
\begin{itemize}
        \item Two distinct horizons ($M>M_{c}$),  
        \item Degenerate horizons ($M=M_{c}$),  
        \item Horizonless spacetime ($M<M_{c}$).  
\end{itemize}

\section{Thermodynamics}
\label{sec3}

\begin{figure}[H]
\begin{minipage}[t]{0.5\textwidth}
        \centering
        \includegraphics[width=\textwidth]{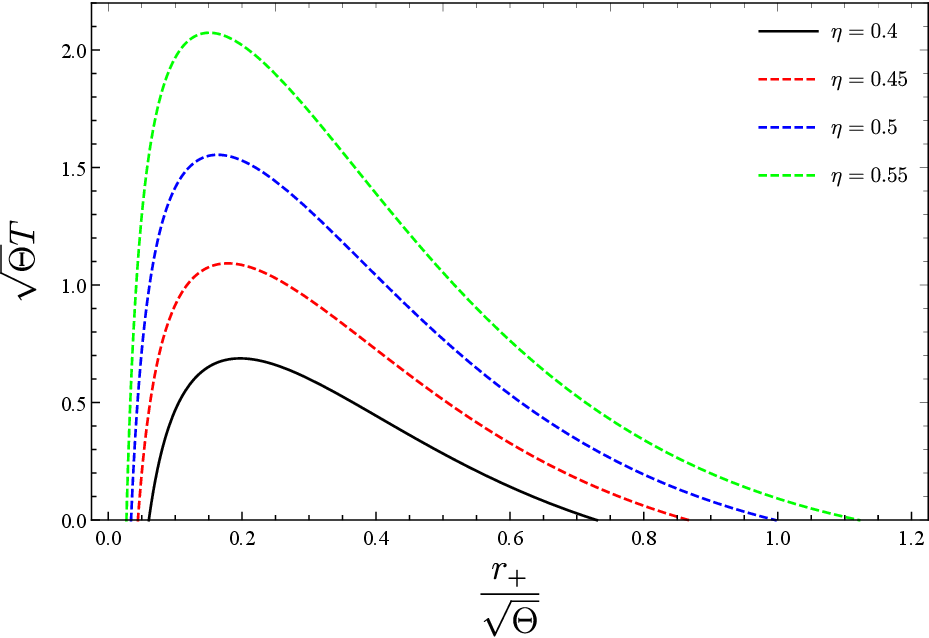}
       \subcaption{ $\frac{\alpha}{\sqrt{\Theta}}=0.2 $}\label{figEC3}
   \end{minipage}%
\begin{minipage}[t]{0.50\textwidth}
        \centering
       \includegraphics[width=\textwidth]{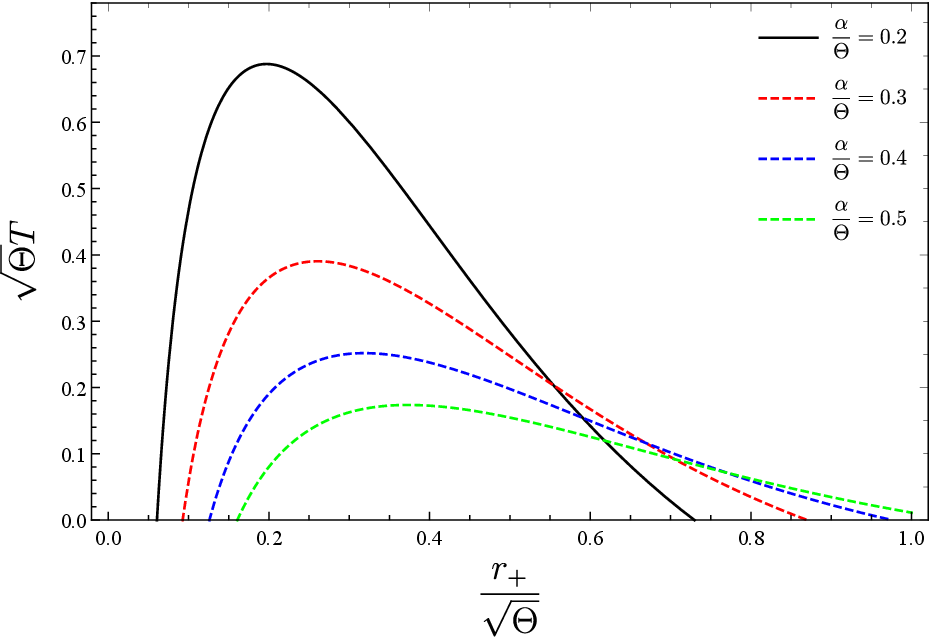}\\
        \subcaption{$\frac{\eta}{\sqrt{\Theta}}=0.4 $}\label{fig:E4}
    \end{minipage}\hfill
\caption{ Hawking temperature vs. $\frac{r_{+}}{\sqrt{\Theta}}$ for different values of $\eta$ and $\alpha$ }
\label{Efigs}
\end{figure}

In this section, we analyze the thermodynamic characteristics of the solution with the metric function given in equation (\ref{sol}). First, we consider the Hawking temperature

\begin{equation}
T=\frac{\kappa }{2\pi },
\end{equation}%
where $\kappa $ represents the surface gravity that can be found from the metric components, namely,
\begin{equation}
\kappa =\frac{1}{2}\left. \frac{\partial }{\partial r}\mathcal{F}\left(
r\right) \right\vert _{r=r_{+}}.  \label{sur}
\end{equation}%
By considering $\left. \mathcal{F}\left( r\right) \right\vert _{r=r_{+}}=0$, we express the black hole mass in terms of the event horizon, the Gauss-Bonnet coupling parameter, the global monopole parameter, and the noncommutative parameter as follows:

\begin{equation}
M=\frac{r_{+}}{2}\left( 1+\frac{\alpha }{r_{+}^{2}}-8\pi \eta ^{2}\right)
\left( 1+\frac{4\sqrt{\Theta }}{\sqrt{\pi }r_{+}}\right) .  \label{mass}
\end{equation}%
By substituting the metric function (\ref{sol}) and the mass expression (\ref{mass}) into Eq. (\ref{sur}), the Hawking temperature can be calculated as: 

\begin{equation}
T=\frac{1}{2\pi r_{+}}\frac{\frac{1}{2}\left( 1+\frac{\alpha }{r_{+}^{2}}%
-8\pi \eta ^{2}\right) \left( 1-\frac{4\sqrt{\Theta }}{\sqrt{\pi }r_{+}}%
\right) -\frac{\alpha }{r_{+}^{2}}}{\left( 1+\frac{2\alpha }{r_{+}^{2}}%
\right) }.  \label{temb}
\end{equation}%
When $\Theta =0$, Eq.(\ref{temb}) simplifies to%
\begin{equation}
T=\frac{1}{4\pi r_{+}}\frac{1-\frac{\alpha }{r_{+}^{2}}-8\pi  \eta ^{2}}{%
1+\frac{2\alpha }{r_{+}^{2}}}.
\end{equation}%
We recover the 4-dimensional Schwarzschild black hole temperature when $\alpha =\eta =0$,

\begin{equation}
T=\frac{1}{4\pi r_{+}}.
\end{equation}%
In Fig.\ref{Efigs}, we present the variation of the Hawking temperature versus $r_{+}/\sqrt{\Theta }$ for different values of the global monopole parameter $\eta $ and GB coupling constant $\alpha $.

Fig. \ref{figEC3} shows the Hawking temperature as a function of $r_{+}/\sqrt{\Theta }$ for different values of the global monopole parameter $\eta $. During black hole evaporation, the Hawking temperature initially increases with $r_{+}/\sqrt{\Theta }$, reaches a maximum at $r_{+}^{c}/\sqrt{\Theta }$, and then rapidly drops to zero. The peak temperature rises with increasing $\eta $, while Fig. \ref{fig:E4} reveals that this maximum temperature is inversely proportional to the Gauss-Bonnet parameter. 
We now use the relation%

\begin{equation}
dM=Td\mathcal{S}
\end{equation}%
which is the first law of black hole thermodynamics. The black hole entropy $\mathcal{S}$ can be obtained from the integration%

\begin{equation}
\mathcal{S}=\int 4\pi r_{+}\left( 1+\frac{2\alpha }{r_{+}^{2}}\right) \frac{%
\frac{1}{2}-4\pi \eta ^{2} -\frac{4\alpha \sqrt{\Theta }}{\sqrt{\pi }%
r_{+}^{3}}-\frac{\alpha }{2r_{+}^{2}}}{1-\frac{\alpha }{r_{+}^{2}}-\left(
8\pi \eta ^{2} +\frac{4\sqrt{\Theta }}{\sqrt{\pi }r_{+}}\left( 1-8\pi
\eta ^{2} +\frac{\alpha }{r_{+}^{2}}\right) \right) }dr_{+}.  \label{24}
\end{equation}%
When $\Theta \rightarrow 0$ , the Einstein-Gauss-Bonnet entropy (\ref{24}) reduces to%

\begin{equation}
\mathcal{S}=\int 2\pi r_{+}\left( 1+\frac{2\alpha }{r_{+}^{2}}\right)
dr_{+}=\pi r_{+}^{2}+4\pi \alpha \log \frac{r_{+}}{r_{0}}.
\end{equation}

Notice that the first term, $S=\pi r_{+}^{2}$, is the standard Bekenstein-Hawking entropy, while the second term represents the effects of the EGB correction. For large $r_{+}$, these corrections become negligible, and the entropy asymptotically approaches the standard Bekenstein-Hawking form.

Now we study local thermodynamic stability by calculating the heat capacity, which informs us about the thermal stability of the black hole under temperature fluctuations. Local stability is determined by the behavior of the heat capacity $C$: a positive value ($C>0$) indicates thermodynamic stability, whereas a negative value ($C<0$) signals instability. To analyze the thermodynamic stability of 4D noncommutative-inspired EGB black holes with global monopoles, we calculate the heat capacity and investigate how the Gauss-Bonnet term influences the thermal behavior. The heat capacity $C$ of the black hole is defined as: 

\begin{equation}
C=\frac{\partial M}{\partial T}=\left( \frac{\partial r_{+}}{\partial T}\right) \left( \frac{\partial M}{\partial r_{+}}\right) ,  \label{29}
\end{equation}

\begin{figure}[H]
\begin{minipage}[t]{0.5\textwidth}
        \centering
        \includegraphics[width=\textwidth]{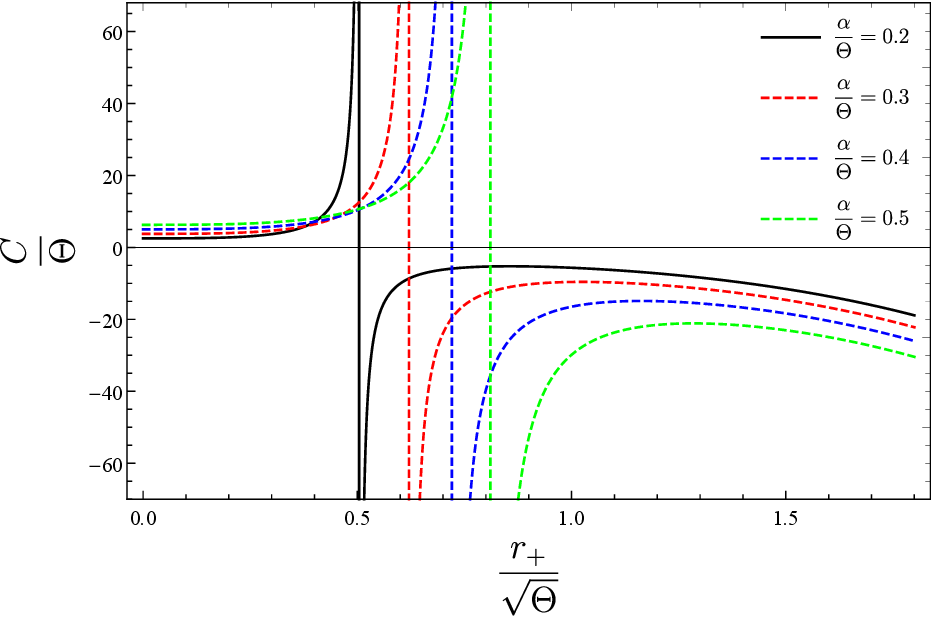}
       \subcaption{ $\frac{\eta}{\sqrt{\Theta}}=0.4$}\label{figheat1}
   \end{minipage}%
\begin{minipage}[t]{0.50\textwidth}
        \centering
       \includegraphics[width=\textwidth]{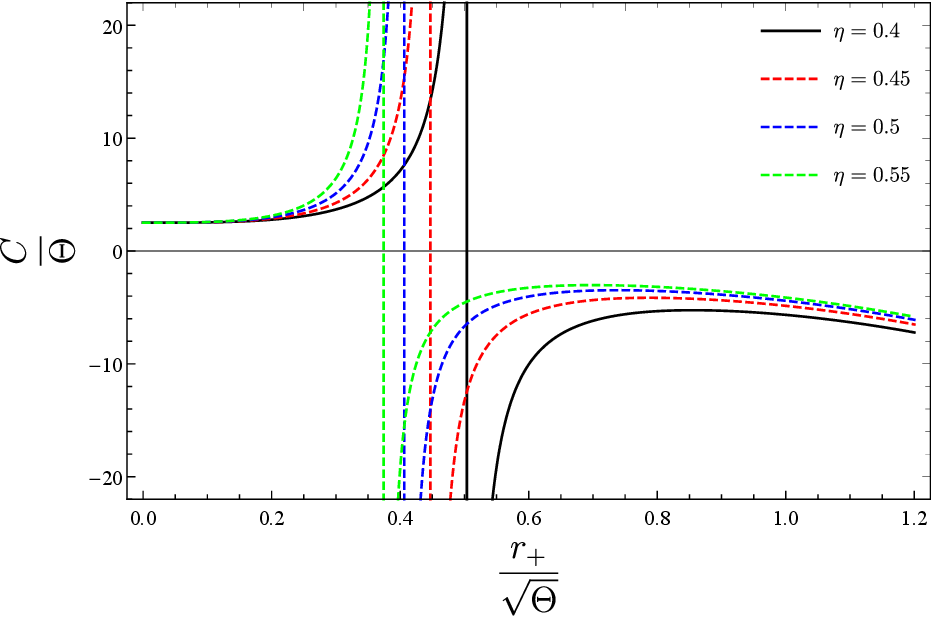}\\
        \subcaption{$\frac{\alpha}{\sqrt{\Theta}}=0.2$}\label{figheat2}
    \end{minipage}\hfill
\caption{The heat capacity vs. $\frac{r_{+}}{\sqrt{\Theta}}$ for different values of $\eta$ and $\alpha$}
\label{capacity}
\end{figure}

By substituting the expressions for mass (Eq. \ref{mass}) and temperature (\ref{temb}) into Eq.(\ref{29}), we obtain the heat capacity for the 4D EGB black holes with global monopoles in the noncommutative spacetime as

\begin{equation}
C=-2\pi r^{2}\frac{\frac{1}{2}\left( 1-\frac{\alpha }{r^{2}}-8\pi  \eta
^{2}\right) \left( 1+\frac{4\sqrt{\Theta }}{\sqrt{\pi }r}\right) -\frac{2%
\sqrt{\Theta }}{\sqrt{\pi }r}\left( 1+\frac{\alpha }{r^{2}}-8\pi  \eta
^{2}\right) }{1+\frac{\frac{2}{r}\sqrt{\frac{\Theta }{\pi }}\left( 1+\frac{%
3\alpha }{r^{2}}-8\pi \eta ^{2}\right) -\frac{3\alpha }{r^{2}}}{\left( 1+%
\frac{2\alpha }{r^{2}}\right) }-\frac{\left( 1+\frac{\alpha }{r^{2}}+\frac{1%
}{2}\left( 1+\frac{\alpha }{r^{2}}-8\pi  \eta ^{2}\right) \left( 1-\frac{4%
\sqrt{\Theta }}{\sqrt{\pi }r}\right) \right) \left( 1-\frac{2\alpha }{r^{2}}%
\right) }{\left( 1+\frac{2\alpha }{r^{2}}\right) ^{2}}}.
\end{equation}%
To investigate the behavior of the heat capacity, we illustrate its
variation in Fig.\ref{capacity} for different values of the global monopole parameter $\eta $ and GB coupling constant $\alpha $.

Figure \ref{figheat1} shows the black hole heat capacity as a function of $\frac{r_{+}}{\sqrt{\Theta}}$ for different values of the GB coupling parameter $\alpha $.

For small values of $ \frac{r_{+}}{\sqrt{\Theta}}$, the heat capacity is positive, indicating thermodynamic stability. As $\frac{r_{+}}{\sqrt{\Theta}}$ increases, each curve exhibits a distinct divergence, marking a second-order phase transition to an unstable thermodynamic phase. Larger values of $\alpha$ shifts the critical point $ r_{+}^c $ to larger radii (rightward), suggesting that GB coupling delays the onset of thermodynamic instability. 
In Fig.\ref{figheat2}, we illustrate the impact of the global monopole parameter $\eta$. Unlike in EGB gravity, higher values of $\eta $  shift the divergence point leftward, toward smaller horizon radii, suggesting that stronger global monopole effects induce phase transitions at smaller radii. Thus, the presence of a global monopole accelerates the onset of thermodynamic stability.

\section{Black Hole Shadow}
\label{sec4}
In this section, we study the shadow behavior of noncommutative Gauss-Bonnet black holes with a global monopole in four-dimensional spacetime. Before exploring shadow geometries in 4-dimensions, we first establish the null geodesic equations of motion. Using the Euler-Lagrange formalism, we derive the equations governing photon trajectories near the black hole horizons, which are given by:%

\begin{equation}
\frac{\partial }{\partial x^{\mu }}\mathcal{L}-\frac{d}{d\lambda }\frac{\partial }{\partial \dot{x}^{\mu }}\mathcal{L}=0.
\end{equation}%

where the dot over the variables stands for the derivative with respect to the affine parameter "$\lambda $". For the metric function given in equation (\ref{sol}), the corresponding Lagrangian can be expressed as:

\begin{equation}
\mathcal{L}=\frac{1}{2}g_{\mu \nu }\dot{x}^{\mu }\dot{x}^{\nu }=\frac{1}{2}%
\left[ -\mathcal{F}\left( r\right) \dot{t}^{2}+\frac{1}{\mathcal{F}\left(
r\right) }\dot{r}^{2}+r^{2}\left( \dot{\theta}^{2}+\sin ^{2}\theta \dot{\varphi}^{2}\right) \right] .
\end{equation}

Next, by employing the lapse function $\mathcal{F}\left( r\right) $, the four-momentum $P_{\mu }$ of a photon is obtained from the Lagrangian as,
\begin{equation}
P_{\mu }=\frac{\partial \mathcal{L}}{\partial \dot{x}^{\mu }}=g_{\mu \nu }\dot{x}^{\nu },
\end{equation}

and thus we have:

\begin{equation}
-P_{t}=\left( 1+\frac{r^{2}}{2\alpha }-\frac{r^{2}}{2\alpha }\sqrt{1+\frac{%
4\alpha }{r^{2}}\left( \frac{2M}{r}+8\pi \eta ^{2}-\frac{8M}{\sqrt{\pi }r^{2}%
}\sqrt{\Theta }\right) }\right) \dot{t}=E,  \label{eeq}
\end{equation}%
\begin{equation}
P_{\varphi }=r^{2}\sin ^{2}\theta \dot{\varphi}=L.  \label{leq}
\end{equation}

Here, $E$ and $L$ correspond to the energy and angular momentum, respectively. Moreover, $P_{t}$ and $P_{\varphi }$ are conserved quantities because the Lagrangian does not depend explicitly on the coordinates $t$ and $\varphi $. The other two geodesic equations can be derived using the Hamilton-Jacobi equation

\begin{equation}
\frac{\partial }{\partial \tau }\mathcal{S}=-\frac{1}{2}g_{\mu \nu }\frac{%
\partial \mathcal{S}}{\partial x^{\mu }}\frac{\partial \mathcal{S}}{\partial x^{\nu }}.  \label{S1}
\end{equation}%

Based on the expressions for the momentum and the Lagrangian derived above, the action $\mathcal{S}$ is assumed to take the following form: 

\begin{equation}
\mathcal{S}=-Et+\mathcal{S}_{r}\left( r\right) +\mathcal{S}_{\theta }\left(\theta \right) +L\varphi .  \label{S2}
\end{equation}%

In this way, the separability of the Hamilton-Jacobi equation leads to%
 
\begin{equation}
r^{2}\mathcal{F}\left( r\right) ^{2}\left( \frac{\partial S_{r}}{\partial r}%
\right) ^{2}=r^{2}E^{2}-\left( \mathcal{K}+L^{2}\right) \mathcal{F}\left( r\right)
,\,\,\,  \label{256}
\end{equation}%

\begin{equation}
\left( \frac{\partial S_{\theta }}{\partial \theta }\right) ^{2}=\mathcal{K}-L^{2}\cot ^{2}\theta ,  \label{28}
\end{equation}%

where $\mathcal{K}$ is called the Carter constant. Using the relation $%
P_{\theta }=\frac{\partial \mathcal{L}}{\partial \dot{\theta}}=\frac{%
\partial \mathcal{S}_{\theta }}{\partial \theta }$, we derive%
\begin{equation}
\frac{\partial \mathcal{S}_{\theta }}{\partial \theta }=r^{2}\frac{\partial\theta }{\partial \tau },  \label{stet}
\end{equation}%

Similarly, from $P_{r}=\frac{\partial \mathcal{S}_{r}}{\partial r}$, we obtain
 
\begin{equation}
\frac{\partial \mathcal{S}_{r}}{\partial r}=r^{2}\frac{\partial r}{\partial\tau }.  \label{ster}
\end{equation}%

Using the relations (\ref{256}), (\ref{28}), (\ref{stet}), and (\ref{ster}), with the definition of the canonically conjugate momentum, we derive the complete set of equations of motion for a massless photon around the black hole

\begin{equation}
r^{2}\left( \frac{\partial r}{\partial \tau }\right) =\pm \sqrt{\mathcal{R}},
\label{530}
\end{equation}%
\begin{equation}
r^{2}\left( \frac{\partial \theta }{\partial \tau }\right) =\pm \sqrt{\Omega},
\end{equation}%

with 
\begin{equation}
\mathcal{R}=E^{2}\left[ r^{4}-r^{2}\left( 1+\frac{r^{2}}{2\alpha }-\frac{%
r^{2}}{2\alpha }\sqrt{1+\frac{4\alpha }{r^{2}}\left( \frac{2M}{r}+8\pi \eta
^{2}-\frac{8M\sqrt{\Theta }}{\sqrt{\pi }r^{2}}\right) }\right) \left( \eta+\zeta ^{2}\right) \right] ,
\end{equation}%

and 
\begin{equation}
\Omega =E^{2}\left[ \eta -\zeta ^{2}\cot ^{2}\theta \right] .
\end{equation}%
Here, $\kappa =\frac{\mathcal{K}}{E^{2}}$ and $\zeta =\frac{L}{E}$ stands for the impact parameters. The radial null geodesic equation (\ref{530}) can be reformulated as:

\begin{equation}
\dot{r}^{2}+V_{\mathrm{eff}}\left( r\right) =0,
\end{equation}%
where $V_{\mathrm{eff}}\left( r\right) $ is the effective radial potential given by
 
\begin{equation}
V_{\mathrm{eff}}\left( r\right) =E^{2}\left[ \frac{\kappa +\zeta ^{2}}{r^{2}}\left( 1+\frac{r^{2}}{2\alpha }-\frac{r^{2}}{2\alpha }\sqrt{1+\frac{4\alpha 
}{r^{2}}\left( \frac{2M}{r}+8\pi \eta ^{2}-\frac{8M\sqrt{\Theta }}{\sqrt{\pi}r^{2}}\right) }\right) -1\right] .  \label{Veff}
\end{equation}%

To determine the photon orbits near the black hole horizon, we impose the following conditions: 

\begin{equation}
\left. V_{\mathrm{eff}}\left( r\right) \right\vert _{r=r_{p}}=\left. \frac{%
\partial }{\partial r}V_{\mathrm{eff}}\left( r\right) \right\vert _{r=r_{\mathrm{p}}}=0,  \label{VL}
\end{equation}%

Applying Eq. (\ref{VL}), the condition $\left. V_{\mathrm{eff}}\left(
r\right) \right\vert _{r=r_{p}}=0$ leads to:%
\begin{equation}
\kappa +\zeta ^{2}=\frac{r_{\mathrm{p}}^{2}}{1+\frac{r_{\mathrm{p}}^{2}}{%
2\alpha }-\frac{r_{\mathrm{p}}^{2}}{2\alpha }\sqrt{1+\frac{4\alpha }{r_{%
\mathrm{p}}^{2}}\left( \frac{2M}{r_{\mathrm{p}}}+8\pi \eta ^{2}-\frac{8M%
\sqrt{\Theta }}{\sqrt{\pi }r_{\mathrm{p}}^{2}}\right) }}.
\end{equation}%
The boundary condition $\left. \frac{\partial }{\partial r}V_{\mathrm{eff}%
}\left( r\right) \right\vert _{r=r_{p}}=0$ leads to: 
\begin{equation}
r_{\mathrm{p}}\mathcal{F}\left( r_{\mathrm{p}}\right) ^{\prime }-2\mathcal{F}%
\left( r_{\mathrm{p}}\right) =0.  \label{psr}
\end{equation}%
By substituting the metric function $\mathcal{F}\left( r\right) $ from Eq. (\ref{sol}) and its first derivative $\mathcal{F}\left( r\right) ^{\prime }$, the above condition simplifies to,%

\begin{equation}
\left( 1-64\pi ^{2}\eta ^{4}\right) r^{4}-48\pi M\eta ^{2}r^{3}+\left(
8\alpha M+96\sqrt{\frac{\Theta }{\pi }}M^{2}\right) r+\left( 256\eta ^{2}M\sqrt{\frac{\Theta }{\pi }}+32\pi \alpha \eta ^{2}-9M^{2}\right)
r^{2}-32\alpha M\sqrt{\frac{\Theta }{\pi }}=0.\label{58}
\end{equation}

The stability of photon orbits at radius $r_{\mathrm{p}}$ is determined by the second derivative of the effective potential:

\begin{itemize}
\item If $\left. \frac{\partial ^{2}}{\partial r^{2}}V_{\mathrm{eff}}\left(
r\right) \right\vert _{r=r_{\mathrm{p}}}>0$, the lightlike geodesics are unstable.

\item If $\left. \frac{\partial ^{2}}{\partial r^{2}}V_{\mathrm{eff}}\left(
r\right) \right\vert _{r=r_{\mathrm{p}}}<0$, the orbits are stable.
\end{itemize}

These unstable photon orbits critically determine the boundary of the black hole's shadow. The analytical solutions of  Eq. (\ref{58}) are cumbersome, hence, we employ numerical methods. Specifically, we compute the photon sphere radius and the corresponding impact parameters by solving Eq. (\ref{58}) numerically for various values of $\alpha ,\eta $ and $\Theta $. The results are summarized in Tables \ref{tab1} and \ref{tab2}.

Table \ref{tab1} presents solutions for the quantity $\kappa +\zeta ^{2}$ for three different values of the GB coupling constant $\alpha $. Our results show that both the photon sphere radius and the impact parameter decrease with the noncommutative parameter $\Theta $. Moreover, for a fixed $\Theta $, these quantities also decrease with $\alpha $.
\begin{table}[H]
\centering%
\begin{tabular}{l|ll|ll|ll}
\hline\hline
& $\alpha =0.1$ &  & $\alpha =0.15$ &  & $\alpha =0.2$ &  \\ \hline
$\Theta $ & $r_{\mathrm{p}}$ & $\kappa +\zeta ^{2}$ & $r_{\mathrm{p}}$ & $\kappa +\zeta ^{2}$
& $r_{\mathrm{p}}$ & $\kappa +\zeta ^{2}$ \\ \hline
0.01 & 3.66812 & 55.7707 & 3.63665 & 55.7454 & 3.60439 & 55.743 \\ 
0.02 & 3.52289 & 52.3487 & 3.48765 & 52.3082 & 3.45129 & 52.2949 \\ 
0.03 & 3.39478 & 49.5723 & 3.35527 & 49.5198 & 3.31418 & 49.5007 \\ 
0.04 & 3.27017 & 47.0932 & 3.22528 & 47.0314 & 3.17804 & 47.0126 \\ 
0.05 & 3.14125 & 44.7659 & 3.08881 & 44.6995 & 3.03255 & 44.6933 \\ 
 \hline
\end{tabular}%
\caption{The values of the photon radius, $r_{\mathrm{p}}$, and $\protect%
\kappa +\protect\zeta ^{2}$ for different values of $\Theta $ with $\protect%
\eta =0.1$ and $M=1$. }
\label{tab1}
\end{table}

\begin{figure}[H]
\begin{minipage}[t]{0.35\textwidth}
        \centering
        \includegraphics[width=\textwidth]{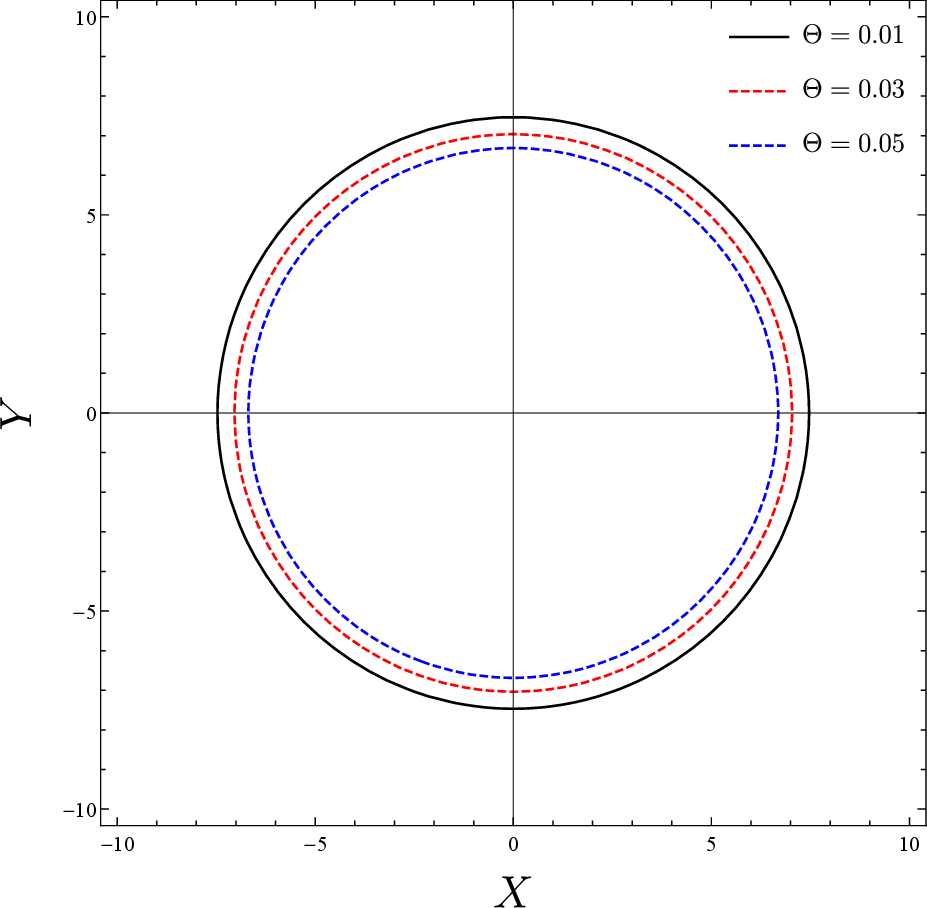}
       \subcaption{ $\alpha=0.1 $}\label{figsh1}
   \end{minipage}%
\begin{minipage}[t]{0.350\textwidth}
        \centering
       \includegraphics[width=\textwidth]{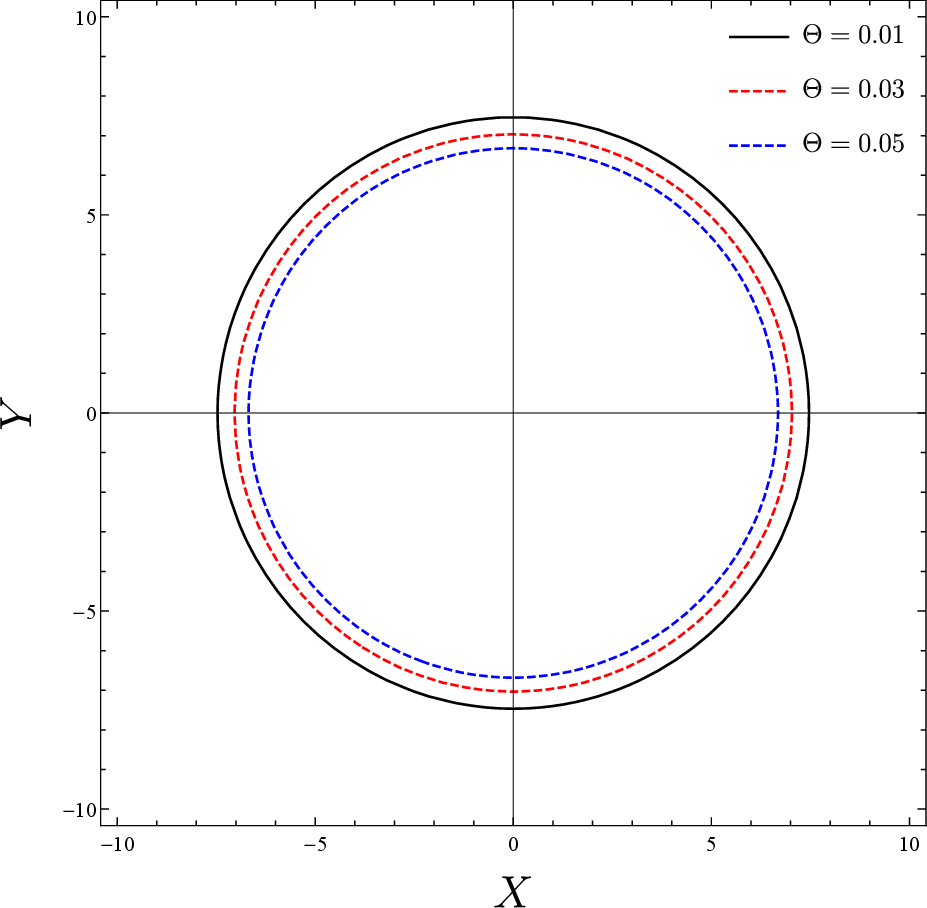}\\
        \subcaption{$\alpha=0.15 $}\label{figsh2}
    \end{minipage}
    \begin{minipage}[t]{0.350\textwidth}
        \centering
       \includegraphics[width=\textwidth]{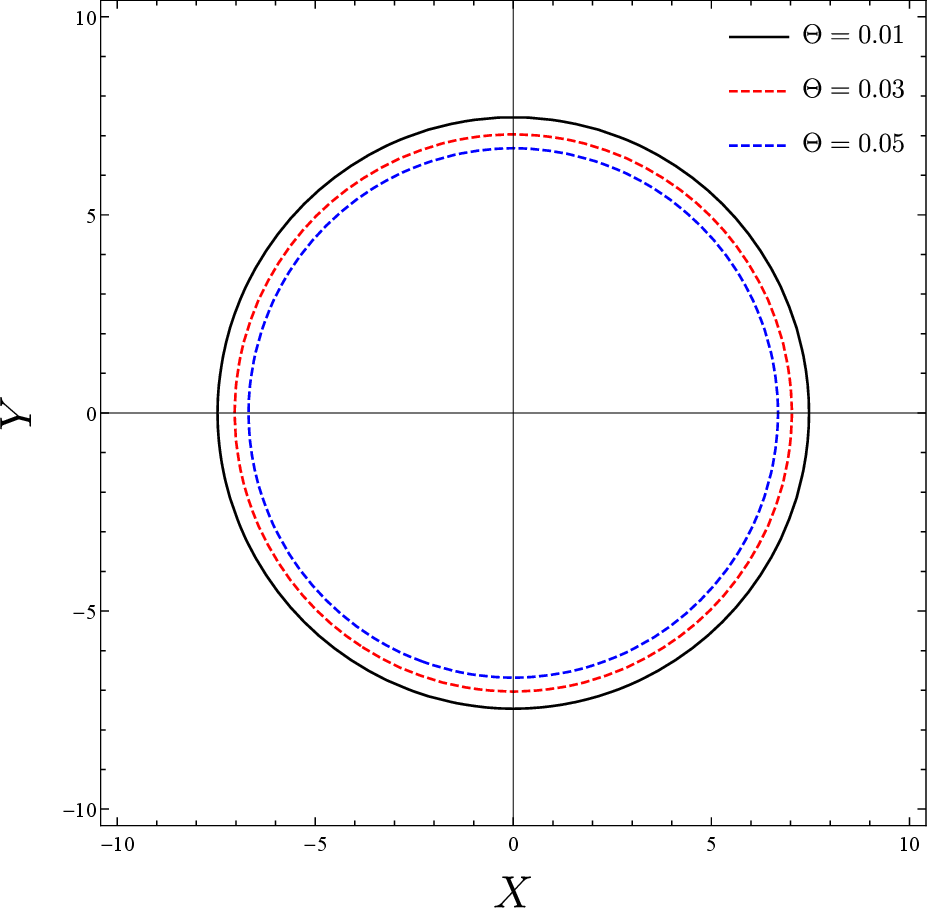}\\
        \subcaption{$\alpha=0.2$}\label{figsh3}
    \end{minipage}\hfill
\caption{Black hole shadow in celestial plane $(X-Y)$ for various values of $\Theta$ and $\alpha$.}
\label{shadow}
\end{figure}

\begin{figure}[H]
\begin{minipage}[t]{0.35\textwidth}
        \centering
        \includegraphics[width=\textwidth]{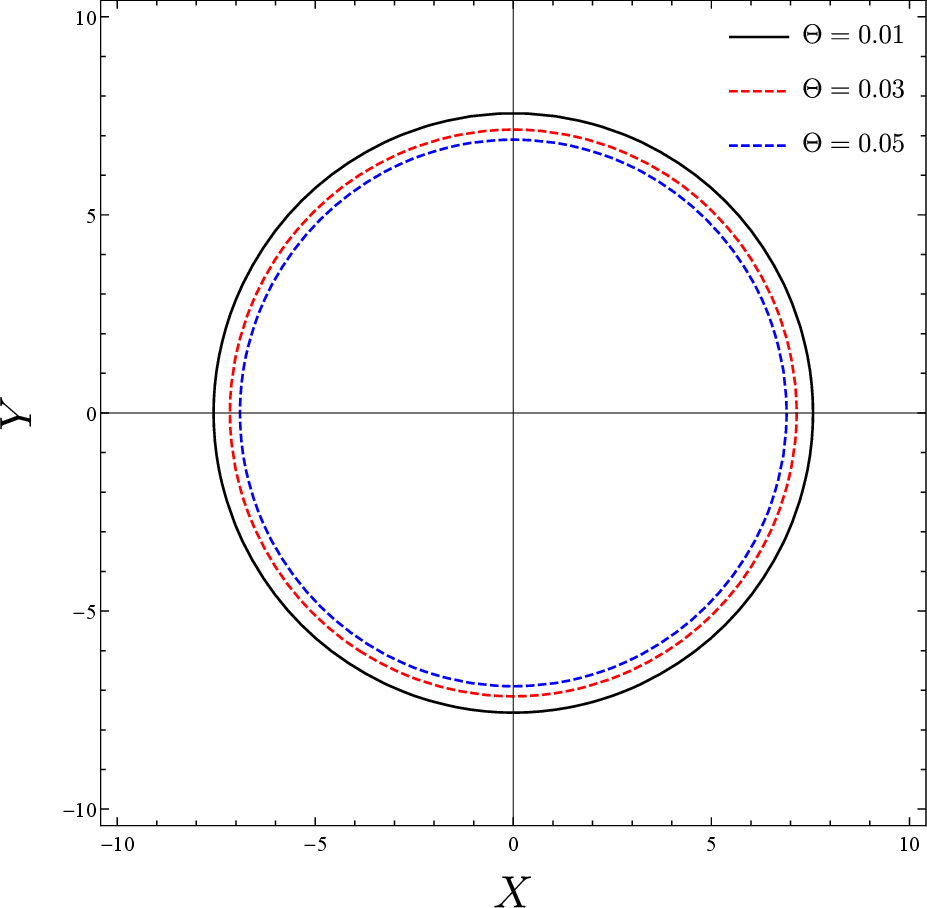}
       \subcaption{ $\eta=0.085 $}\label{figsh4}
   \end{minipage}%
\begin{minipage}[t]{0.350\textwidth}
        \centering
       \includegraphics[width=\textwidth]{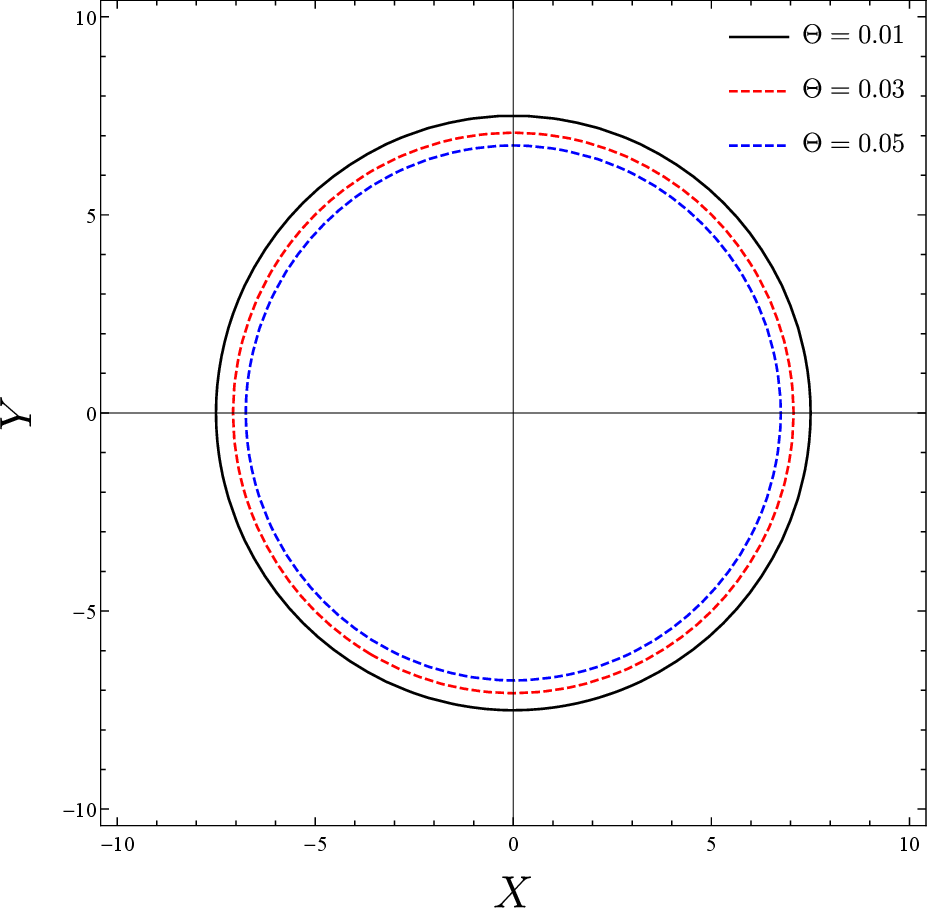}\\
        \subcaption{$\eta=0.09 $}\label{figsh5}
    \end{minipage}
    \begin{minipage}[t]{0.350\textwidth}
        \centering
       \includegraphics[width=\textwidth]{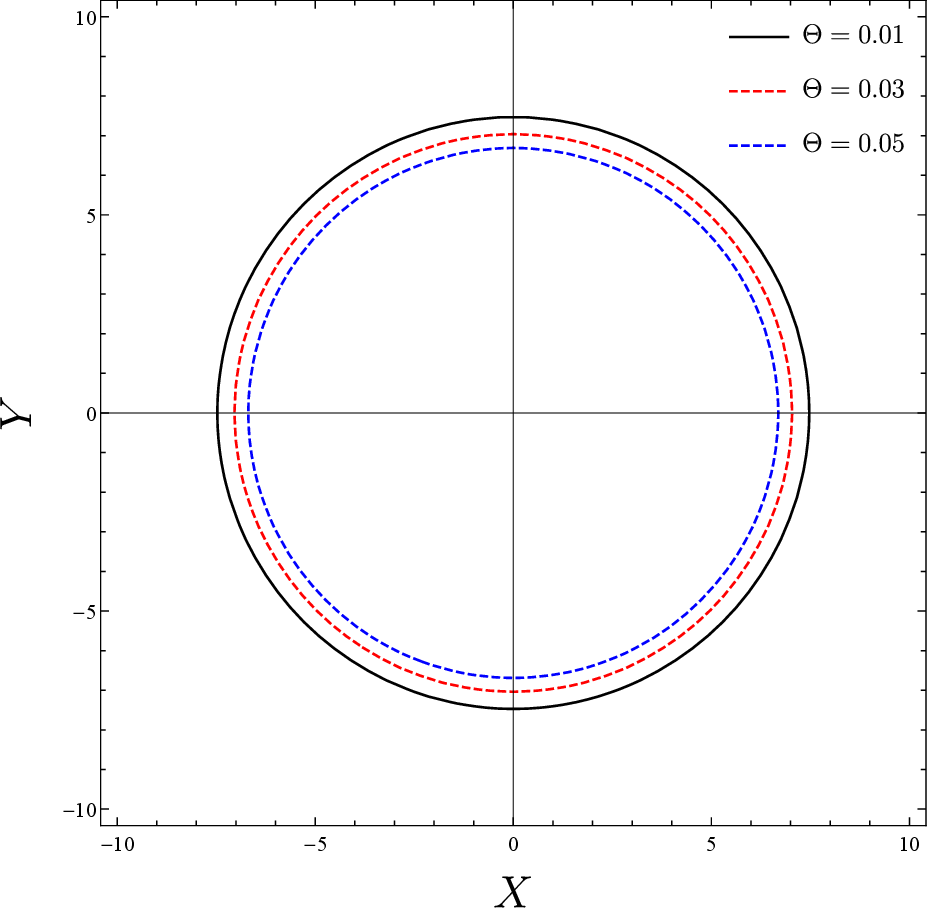}\\
        \subcaption{$\eta=0.095 $}\label{figsh6}
    \end{minipage}\hfill
\caption{Black hole shadow in celestial plane $(X-Y)$ for various values of $\Theta$ and $\eta$.}
\label{figshadow}
\end{figure}

\begin{table}[H]
\centering%
\begin{tabular}{l|ll|ll|ll}
\hline\hline
& $\eta =0.085$ &  & $\eta =0.09$ &  & $\eta =0.095$ &  \\ \hline
$\Theta $ & $r_{\mathrm{p}}$ & $\kappa +\zeta ^{2}$ & $r_{\mathrm{p}}$ & $\kappa +\zeta ^{2}$
& $r_{\mathrm{p}}$ & $\kappa +\zeta ^{2}$ \\ \hline
0.01 & 3.30429 & 57.2432 & 3.41274 & 56.3222 & 3.53345 & 55.8295 \\ 
0.02 & 3.13953 & 53.8136 & 3.25456 & 52.8425 & 3.38179 & 52.3591 \\ 
0.03 & 2.98687 & 51.171 & 3.11055 & 50.0739 & 3.24595 & 49.5582 \\ 
0.04 & 2.82723 & 49.0487 & 2.96409 & 47.6846 & 3.11107 & 47.0771 \\ 
0.05 & 2.63905 & 47.6364 & 2.80093 & 45.6094 & 2.96694 & 44.7833 \\ \hline
\end{tabular}%
\caption{The values of the photon radius, $r_{\mathrm{p}}$, and $\protect%
\kappa +\protect\zeta ^{2}$ for different values of $\Theta $ with $\protect%
\alpha =0.1$ and $M=1$.}
\label{tab2}
\end{table}
On the other hand, the shadow of a black hole and its associated photon orbits can be analyzed using the framework of geometrical optics. The apparent shape of the black hole is governed by the shadow's boundary. To visualize the black hole shadow, it is convenient to employ celestial coordinates ($X$,$Y$). Following Ref. \cite{Synge}, the celestial coordinates can be defined as:
\begin{align}
&X=\lim_{r_{\mathrm{o}}\rightarrow \infty }\left( -r_{\mathrm{o}}^{2}\sin \theta _{\mathrm{o}}\frac{d\varphi }{dr}\right) ,  \label{XX} \\
&Y=\lim_{r_{\mathrm{o}}\rightarrow \infty }\left( r_{\mathrm{o}}^{2}\frac{d\theta }{dr}\right) .  \label{YY}
\end{align}
Here, $r_{\mathrm{o}}$ represents the distance between the black hole and a distant observer, while $\theta _{\mathrm{o}}$ denotes the angular position of the observer relative to the black hole's equatorial plane. For a static observer at a large distance, i.e. at $r_{\mathrm{o}}\rightarrow \infty $ in the equatorial plane $\theta _{\mathrm{o}}=\pi /2$, the celestial coordinates simplify to%

\begin{equation}
X^{2}+Y^{2}=\kappa +\zeta ^{2}=R_{\mathrm{sh}}^{2},
\end{equation}

where $R_{\mathrm{sh}}$ is the radius of the shadow. Finally, we illustrate the dependence of the shadow radius, $R_{\mathrm{sh}}$, on the parameters $\alpha $, and $\eta $ in Figures \ref{shadow}-\ref{figshadow}. Our results demonstrate that these parameters significantly influence the black hole shadow.
 
Next, we investigate the energy emission rate of the 4D EGB black hole with global monopoles inspired by noncommutative geometry. The expression of energy emission rate reads
\begin{equation}
\frac{d^{2}Z\left( \omega \right) }{dtd\omega }=\frac{2\pi ^{2}\sigma _{\lim
}}{e^{\frac{\omega }{T}}-1}\omega ^{3}.
\end{equation}

Here, $\omega $ represents the frequency of the photon, and the parameter $\sigma _{\lim }$ closely approximates to the geometrical cross section of the photon sphere:
\begin{equation}
\sigma _{\lim }\simeq \pi R_{\mathrm{sh}}^{2}.  \label{252}
\end{equation}

Using (\ref{252}), we obtain the expression of the energy emission rate in
the presence of DE in 4D as

\begin{equation}
\frac{d^{2}Z\left( \omega \right) }{dtd\omega }=\frac{2\pi ^{3}R_{\mathrm{sh}%
}^{2}}{e^{\frac{\omega }{T}}-1}\omega ^{3}.
\end{equation}

\begin{figure}[H]
\begin{minipage}[t]{0.35\textwidth}
        \centering
        \includegraphics[width=\textwidth]{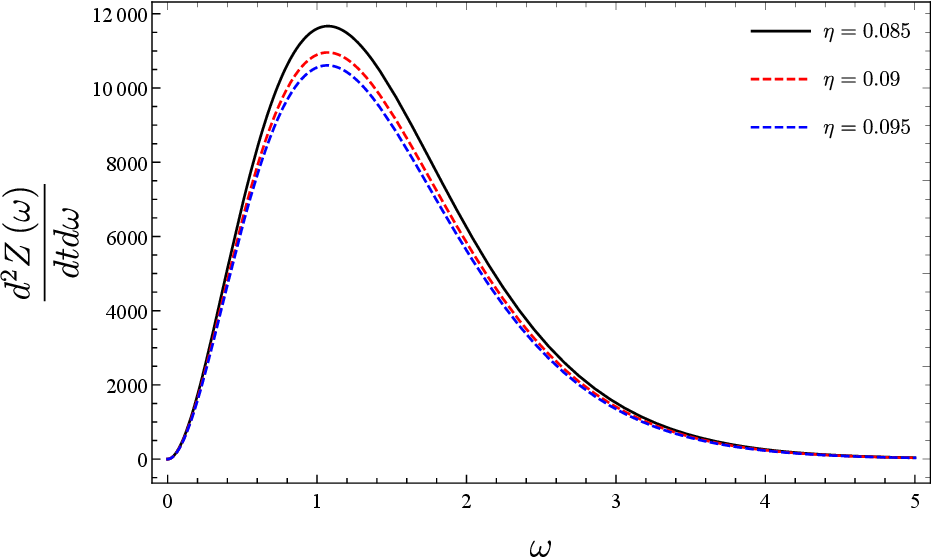}
       \subcaption{ $\alpha=0.1$, $\Theta=0.4$}\label{figem1}
   \end{minipage}%
\begin{minipage}[t]{0.350\textwidth}
        \centering
       \includegraphics[width=\textwidth]{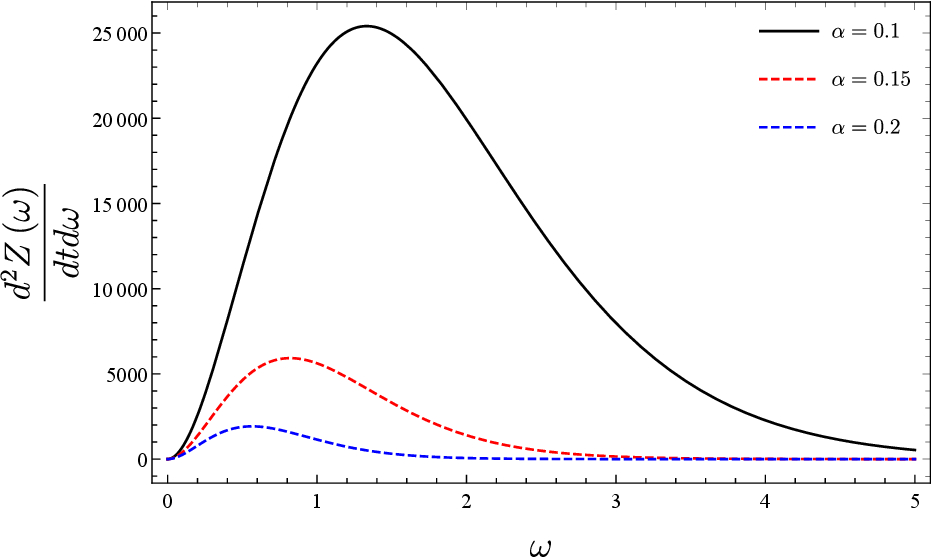}\\
        \subcaption{$\eta=0.1$, $\Theta=0.02$}\label{figem2}
    \end{minipage}
    \begin{minipage}[t]{0.350\textwidth}
        \centering
       \includegraphics[width=\textwidth]{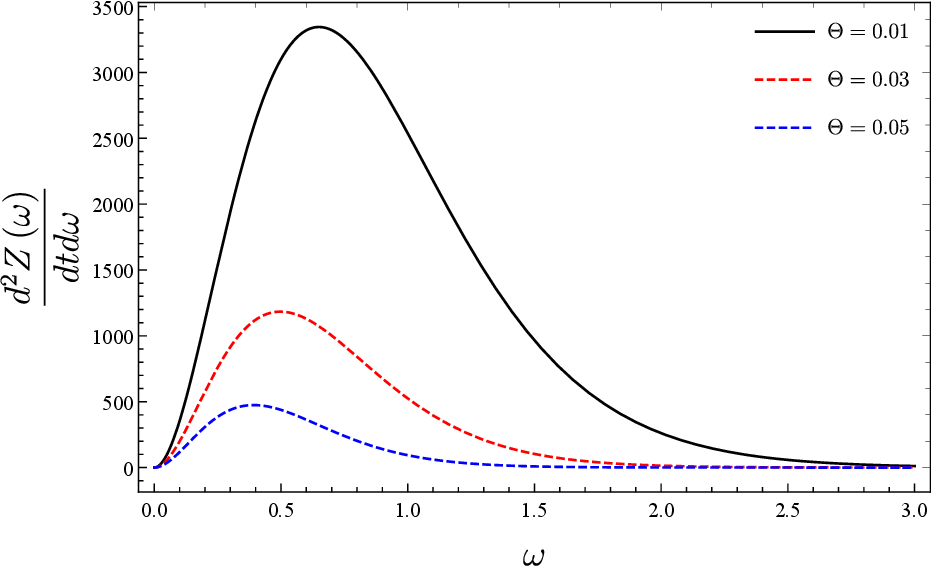}\\
        \subcaption{$\eta=0.1 $, $\alpha=0.2$}\label{figem3}
    \end{minipage}\hfill
\caption{Energy emission rate behavior vs. the  frequency for various values of $\alpha $, $\Theta $ and $\eta $.}
\label{figemission}
\end{figure}

The energy emission rate is shown in Fig.\ref{figemission} as a function of $\omega $ for various values of $\alpha$ ,$\Theta $ and $\eta $. We observe that the energy emission rate decreases with increasing $\alpha$, $\Theta $ and $\eta$.
\section{Quasinormal Modes}
\label{sec5}
When a black hole is perturbed, it undergoes oscillations. If the black hole is stable, the amplitude of these oscillations gradually decays over time; in contrast, if the black hole is unstable, the amplitude grows. The frequency associated with these quasinormal modes (QNMs) is a complex quantity: the real part determines the oscillation frequency, while the imaginary part describes the damping caused by gravitational wave emission. To begin our investigation, we consider the scalar field due to its relatively simple mathematical description. Specifically,  a massless scalar field, $\Psi $, is described by the Klein-Gordon equation, which in the
presence of a background metric $g_{\mu \nu }$, takes the form:%
\begin{equation}
\frac{1}{\sqrt{-g}}\partial _{\mu }\left( \sqrt{-g}g^{\mu \nu }\partial
_{\nu }\Psi \right) =0.
\end{equation}%
The scalar field function $\Psi $ admits a spherical harmonic decomposition:%
\begin{equation}
\Psi =\frac{1}{r}\sum_{\ell ,m}e^{-i\omega t}\mathcal{Y}_{\ell ,m}\left(\theta ,\varphi \right) u_{n,\ell }\left( r\right) ,
\end{equation}

where $\ell $ and $m$ represent the angular and magnetic quantum numbers, respectively, while $\omega $ denotes the characteristic oscillation frequency of the scalar field. The function $\mathcal{Y}_{\ell ,m}\left(\theta ,\varphi \right) $ denotes the standard spherical harmonics and $u_{n,\ell }\left( r\right) $ is the radial function that satisfies the following equation:%

\begin{equation}
\frac{d^{2}}{dr_{\star }^{2}}u_{n,\ell }\left( r\right) +\left( \omega ^{2}-%
\mathcal{V}_{\mathrm{S}}\left( r\right) \right) u_{n,\ell }\left( r\right)=0,  \label{314}
\end{equation}%

where%
\begin{equation}
\mathcal{V}_{\mathrm{S}}\left( r\right) =\mathcal{F}\left( r\right) \left( 
\frac{\ell \left( \ell +1\right) }{r^{2}}+\frac{\mathcal{F}\left( r\right)^{\prime }}{r}\right) ,
\end{equation}

is the effective potential, and $r_{\star }$ is the tortoise coordinates which is defined by the transformation, $dr_{\star }=\frac{dr}{\mathcal{F}\left( r\right) }$. The QNMs are defined as solutions of the radial wave equation (\ref{314}) that fulfill the following boundary conditions:

\begin{itemize}
\item Purely ingoing waves as $r_{\star }\rightarrow -\infty ,$%
\begin{equation}
u_{n,\ell }\left( r\right) \simeq e^{-i\omega r_{\star }}.
\end{equation}

\item Purely outgoing waves as $r_{\star }\rightarrow +\infty ,$%
\begin{equation}
u_{n,\ell }\left( r\right) \simeq e^{i\omega r_{\star }}.
\end{equation}
\end{itemize}

The oscillation frequency can be decomposed into its real and imaginary components as $\omega =\omega _{\mathrm{Re}}-i\omega _{\mathrm{Im}}$.

The QNM frequencies are usually determined numerically using various methods, each suited to specific spacetime geometries and perturbation types. Among these methods are the WKB approximation \cite{Schutz,Will,Sai}, the P\"{o}schl-Teller potential method \cite{Teller}, time-domain integration \cite{Carsten}, Leaver's continued fraction technique \cite{Leaver,EdwardW}, the Bernstein spectral method \cite{Sean}, and the correspondence between QNMs and black hole shadow radii \cite%
{Cardoso,Stefanov,KJusufi}.

\subsection{WKB Approximation}

The main motivation for employing the WKB method lies in the fact
that perturbative equations for particles closely resemble the 1D Schr\"{o}dinger equation describing a potential barrier. A detailed treatment of the WKB approach, including boundary condition matching for deriving QNMs\ is provided in Refs. \cite{Schutz,Will,Sai}, with higher-order extensions discussed in Refs. \cite{Konoplya,Froeman,Andersson,Linn,ESimone,Zhidenko2010,Zhidenkophy,ZinhailoClass}.

The QNM frequencies can be calculated using the following expression:

\begin{equation}
i\frac{\left( \omega -V_{0}\right) }{\sqrt{-2V_{0}''}}-\sum_{i=2}^{N}\Lambda
_{i}=n+\frac{1}{2}.  \label{qnmf}
\end{equation}%
Here, $V_{0}$ and $V_{0}''$ represent the effective potential and its second derivative with respect to $r^{\ast }$ at its maximum point "$r_{0}^{\ast }$", respectively. The term $\Lambda _{i}$ corresponds to a constant coefficient arising from higher-order WKB corrections, while $n=0,1,2,...$ denotes the overtone number. It is worth mentioning that Eq.(\ref{qnmf})  involves nontrivial functions of physical parameters; therefore, we compute the QNM frequencies for various values of these parameters. However, given that the WKB formula provides the most accurate results when $\ell >n$ \cite{Stuchlik}, we restrict our analysis to scalar field modes that satisfy this condition.

We calculate the QNM frequencies using the 6th-order WKB approximation for a massless scalar field with the specified parameters. The results are summarized in Tables \ref{tab:week4}-\ref{tab:week5}.
\begin{table}[H]
\centering%
\begin{tabular}{|l|l|lll|}
\hline\hline
\rowcolor{lightgray} $\ell $ & $n$ & $\Theta =0.02$ & $\Theta =0.04$ & $\Theta =0.06$ \\
 \hline
2 & 0 & 0.345495 -i0.0544776 & 0.364307 -i0.0538304 & 0.383161 -i0.0520348
\\ 
& 1 & 0.326593 -i0.166959 & 0.347749 -i0.165442 & 0.369041 -i0.160032 \\ 
\hline
3 & 0 & 0.483812 -i0.0543121 & 0.510237 -i0.0536724 & 0.536829 -i0.0518667
\\ 
& 1 & 0.473543 -i0.166682 & 0.501312 -i0.164509 & 0.528994 -i0.158565 \\ 
& 2 & 0.437434 -i0.272117 & 0.468979 -i0.271564 & 0.501529 - 0.264412 \\ 
\hline
4 & 0 & 0.656128 -i0.0536051 & 0.656128 -i0.0536051 & 0.690418 -i0.0517954
\\ 
& 1 & 0.650238 -i0.163236 & 0.650238 -i0.163236 & 0.685093 -i0.157376 \\ 
& 2 & 0.629831 -i0.275145 & 0.629831 -i0.275145 & 0.667988 -i0.265598  \\ 
& 3 & 0.588576 -i0.374436 & 0.588576 -i0.374436  & 0.631975 -i0.366437  \\ 
\hline
5 & 0 & 0.760438 -i0.0542096 & 0.802009 -i0.0535711 & 0.843979 -i0.0517593
\\ 
& 1 & 0.755486 -i0.164552 & 0.797643 -i0.162425  & 0.839962 -i0.156671  \\ 
& 2 & 0.739593 -i0.27791 & 0.783958 -i0.274342 & 0.828303 -i0.264339  \\ 
& 3 & 0.704953 -i0.385891 & 0.75345 -i0.383403  & 0.802798 -i0.371401 \\ 
& 4 & 0.657416 -i0.47342  & 0.708095 -i0.475787 & 0.762024 -i0.467237  \\ 
\hline\hline
\end{tabular}%
\caption{QNMs for different values of $\Theta$, $\ell$ and $n$ with $M=1$, $\alpha=0.1$ and $\eta =0.1$.}
\label{tab:week4}
\end{table}

\begin{table}[H]
\centering%
\begin{tabular}{|l|l|lll|}
\hline\hline
\rowcolor{lightgray} $\ell $ & $n$ & $\alpha =0.1$ & $\alpha =0.15$ & $%
\alpha =0.2$ \\ \hline
2 & 0 & 0.334622- i0.0544668 & 0.336232- i0.0541024 & 0.337882- i0.0537152
\\ 
& 1 & 0.314469- i0.16655  & 0.316645- i0.165577 & 0.318857- i0.164522 \\ 
\hline
3 & 0 & 0.468613- i0.054301 & 0.470857- i0.0539351 & 0.473172- i0.0535479 \\ 
& 1 & 0.457614- i0.166717 & 0.460175- i0.165566 & 0.46282- i0.164352 \\ 
& 2 & 0.419896- i0.270451  & 0.423267- i0.269225 & 0.426753- i0.267911 \\ 
\hline
4 & 0 & 0.602591- i0.054232  & 0.605467- i0.053865 & 0.608448- i0.0534779 \\ 
& 1 & 0.595427- i0.165493  & 0.59847- i0.164325 & 0.601673- i0.163106 \\ 
& 2 & 0.570288- i0.27759 & 0.574025- i0.275822 & 0.577992- i0.273993  \\ 
& 3 & 0.480625- i0.436815 & 0.529095- i0.369059 & 0.533785 - 0.367549 \\ 
\hline
5 & 0 & 0.736554- i0.0541965 & 0.740079- i0.0538303 & 0.743713- i0.0534422
\\ 
& 1 & 0.731283- i0.164606 & 0.734941- i0.163457  & 0.738713- i0.162242 \\ 
& 2 & 0.714252- i0.277915 & 0.718412- i0.276004 & 0.722706- i0.273973 \\ 
& 3 & 0.677776- i0.384412 & 0.682816- i0.382325 & 0.722706- i0.273973 \\ 
& 4 & 0.62983- i0.468932  & 0.63528- i0.467373 & 0.641008- i0.465643 \\ 
\hline\hline
\end{tabular}%
\caption{QNMs for different values of $\alpha$, $\ell$ and $n$ with $M=1$, $\Theta=0.01$ and $\eta =0.1$.}
\label{tab:week3}
\end{table}
\begin{table}[tbph]
\centering%
\begin{tabular}{|l|l|lll|}
\hline\hline
\rowcolor{lightgray} $\ell $ & $n$ & $\eta =0.07$ & $\eta =0.08$ & $\eta
=0.09$ \\ \hline
2 & 0 & 0.431234 -i0.074478  & 0.401739 -i0.0682867  & 0.369403 -i0.0615767
\\ 
& 1 & 0.400985 -i0.22671 & 0.374676 -i0.20815  & 0.345738 -i0.187988 \\ 
\hline
3 & 0 & 0.602915 -i0.0742155 & 0.561955 -i0.0680567  & 0.516994 -i0.0613772
\\ 
& 1 & 0.586244 -i0.228111 & 0.547102 -i0.209127  & 0.504015 -i0.188528 \\ 
& 2 & 0.531144 -i0.364982 & 0.497444 -i0.335926 & 0.460134 -i0.304232 \\ 
\hline
4 & 0 & 0.774768 -i0.0741053 & 0.722271 -i0.0679593 & 0.664643 -i0.0612949 
\\ 
& 1 & 0.764006 -i0.226504  & 0.712644 -i0.207626 & 0.656223 -i0.187165 \\ 
& 2 & 0.725676 -i0.378118 & 0.678487 -i0.347107 & 0.6265 -i0.3134 \\ 
& 3 & 0.661625 -i0.497661 & 0.620064 -i0.458606 & 0.574265 -i0.41599 \\ 
\hline
5 & 0 & 0.946692 -i0.0740489 & 0.882633 -i0.0679099 & 0.812296 -i0.0612521
\\ 
& 1 & 0.938845 -i0.225211 & 0.875595 -i0.20646 & 0.806101 -i0.18613  \\ 
& 2 & 0.912662 -i0.379928  & 0.852321 -i0.348394  & 0.785873 -i0.314162 \\ 
& 3 & 0.858377 -i0.520916 & 0.803592 -i0.478903  & 0.743108 -i0.433116 \\ 
& 4 & 0.793042 -i0.628339 & 0.743488 -i0.579411 & 0.688797 -i0.52599 \\ 
\hline\hline
\end{tabular}%
\caption{QNMs for different values of $\eta$, $\ell$ and $n$ with $M=1$, $\alpha=0.01$ and $\Theta =0.01$.}
\label{tab:week5}
\end{table}
The QNM frequencies for selected values of $n$, $\ell$, and $\Theta$ are presented in Table \ref{tab:week4}. We observe that the imaginary parts of these frequencies are negative for all branches, indicating a stable propagation of scalar fields in this background. This stability arises from the presence of a positive potential barrier outside the event horizon. For a fixed value of $\Theta$, the real part of the frequency $\omega_{\mathrm{Re}}$ increases with increasing angular momentum number $\ell$. However, for a given $\ell$, an increase in the coupling constant $\Theta$ leads to a decrease in $\omega_{\mathrm{Re}}$ for each overtone number $n$.

\begin{figure}[H]
\begin{minipage}[t]{0.5\textwidth}
        \centering
        \includegraphics[width=\textwidth]{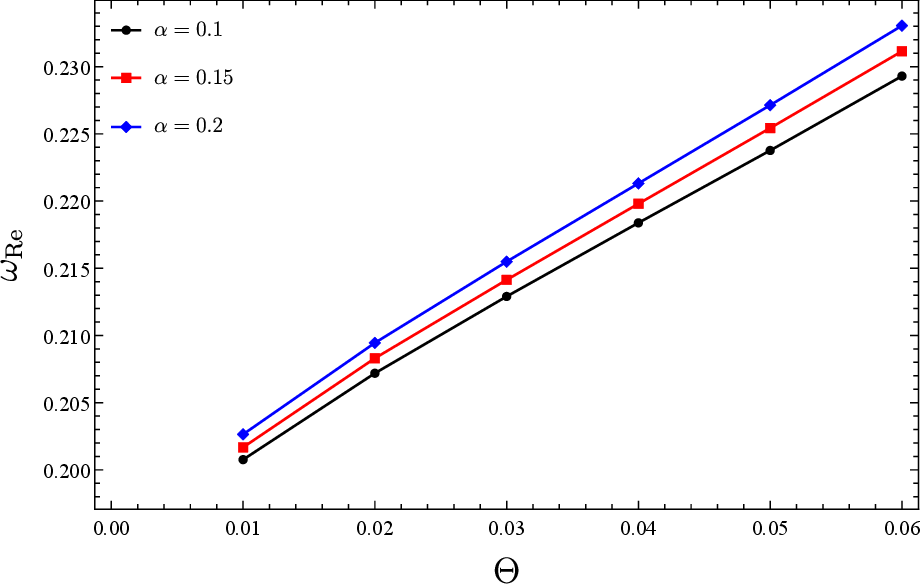}
       \label{figq1}
   \end{minipage}%
    \begin{minipage}[t]{0.50\textwidth}
        \centering
       \includegraphics[width=\textwidth]{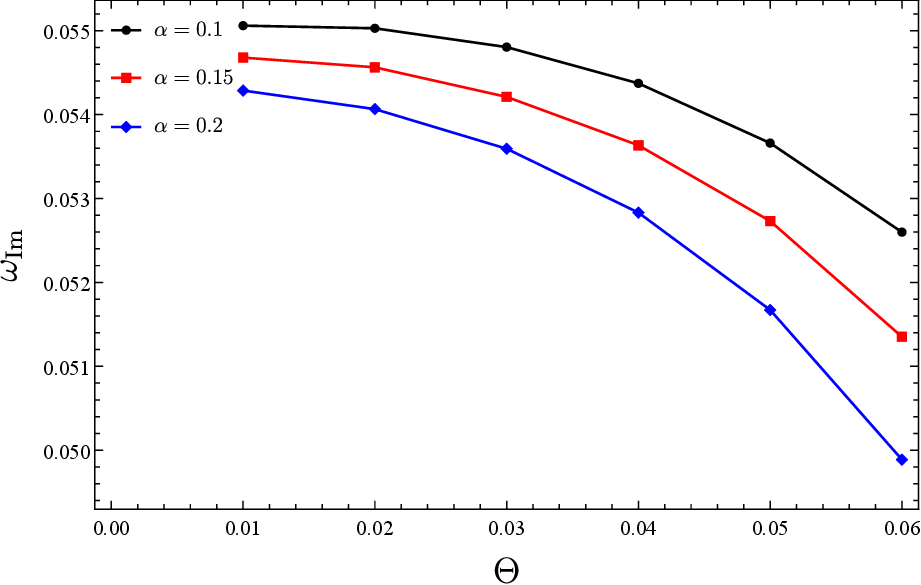}\\
        \label{figq2}
    \end{minipage}\hfill
\caption{Fundamental QNMs of black holes in a scalar field for different values of $\alpha$ and $\eta=0.1$, $\Theta=0.01$, $M=1$}
\label{figqnm1}
\end{figure}

In Fig.\ref{figqnm1}, we plot the real and imaginary components of the quasinormal frequencies as functions of $\Theta$ for different values of the GB coupling constant $\alpha$. We observe that as $\Theta$ increases, the real part of the frequency grows monotonically, while the imaginary part decreases. Additionally, Fig.\ref{figqnm1}  also shows that for a fixed value of $\Theta$, the real part increases with $\alpha$, whereas the imaginary part decreases.

\begin{figure}[H]
\begin{minipage}[t]{0.5\textwidth}
        \centering
        \includegraphics[width=\textwidth]{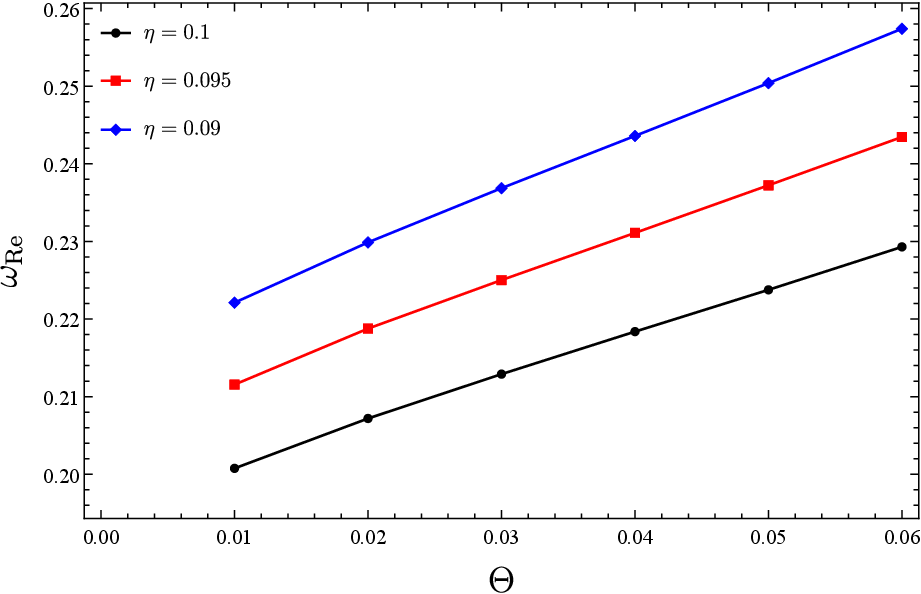}
       \label{figq3}
   \end{minipage}%
    \begin{minipage}[t]{0.50\textwidth}
        \centering
       \includegraphics[width=\textwidth]{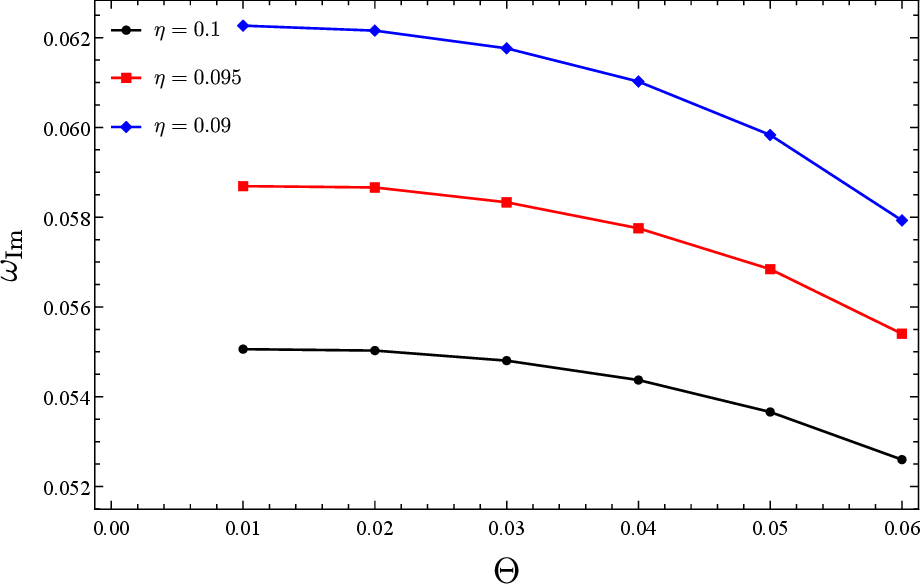}
       \label{figq4}
    \end{minipage}\hfill
\caption{Fundamental QNMs of black holes in a scalar field for different values of $\eta$ and $\alpha=0.1$, $\Theta=0.01$, $M=1$ }
\label{figqnm2}
\end{figure}

Figure \ref{figqnm2} shows the dependence of both the real ($\omega_{\mathrm{Re}}$) and imaginary ($\omega_{\mathrm{Im}}$) parts of the fundamental QNMs on the parameter $\Theta$ for various values of the global monopole parameter $\eta$. The plots reveal that both $\omega_{\mathrm{Re}}$ and $\omega_{\mathrm{Im}}$ decrease monotonically with increasing $\eta$.

\subsection{Unstable null geodesics and quasinormal modes}
The detection of black holes through gravitational waves and shadow imaging makes it interesting to explore the relationship between QNMs and the geodesics that describe light propagation near black holes. An important remark in this field was presented in \cite{Cardoso}, which noted that the properties of unstable circular null geodesics around stationary, spherically symmetric, and asymptotically flat black holes are closely linked to the QNMs emitted by these black holes in the regime where $\ell >>n$.  Thus, in the eikonal regime, the radius of the black hole's photon sphere $r_{\mathrm{p}}$ can be directly linked to its QNMs through the analytical relation
\begin{equation}
\omega _{\mathrm{sh}} =\frac{\sqrt{\mathcal{F}\left( r_{\mathrm{p}}\right) }}{r_{\mathrm{p}}} {\ell}-i\lambda \left(n+\frac{1}{2}\right),\label{omegasha}
\end{equation}%
where
\begin{equation}
\lambda =\sqrt{\frac{\left( 2\mathcal{F}\left( r_{\mathrm{p}}\right) -r_{\mathrm{p}}^{2}\mathcal{F}^{\prime \prime }\left( r_{\mathrm{p}}\right) \right) \mathcal{F}\left( r_{\mathrm{p}}\right) }{2r_{\mathrm{p}}^{2}}},
\end{equation}
is the Lyapunov exponent.

The relation (\ref{omegasha}) can be understood by considering that gravitational waves, in the eikonal limit, behave as massless particles traveling along the last unstable null orbit before escaping to infinity. However, this correspondence is violated for gravitational perturbations or non-minimally coupled fields. Additionally, Eq. (\ref{omegasha}) loses accuracy for small values of the multipole number. 

Using Eq. (\ref{omegasha}), we calculate the eikonal QNM frequencies and present the results in Table \ref{tab5}. These values are compared with those derived from the 6th-order WKB approximation. 
\begin{table}[H]
\centering%
\begin{tabular}{l|ll||ll}
\hline\hline
& $\Theta =0.01$ &  & $\Theta =0.03$   \\ \hline
$\ell $ & $\omega_{\mathrm{sh}}$ & $\omega_{\mathrm{WKB}}$ & $\omega_{\mathrm{sh}}$ & $\omega_{\mathrm{WKB}}$ \\ \hline
50 & 6.69525 -i0.162716 & 6.76624 -i0.162332 & 7.1015 -i0.162917 & 7.1798 -i0.161759  \\ 
70 & 9.37336 -i0.162716 & 9.44442 -i0.162352  & 9.94211 -i0.162917 & 10.0219 -i0.161777  \\ 
100 & 13.3905 -i0.162716 & 13.4622 -i0.162363 & 14.203 -i0.162917 & 14.2854 -i0.161787  \\ 
150 & 20.0858 -i0.162716 & 20.1588 -i0.162369 & 21.3045 -i0.162917 & 21.3917 -i0.161792 \\ 
200 & 26.781 - 0.162716 & 26.8557 -i0.162371 & 28.406 -i0.162917  & 28.4982 -i0.161794  \\ 
250 & 33.4763 -i0.162716 & 33.5526 -i0.162372 & 35.5075 -i0.162917 & 35.6047 -i0.161794   \\
300 & 40.1715 -i0.162716 & 40.2495 -i0.162373  & 42.609 -i0.162917 & 42.7113 -i0.161795  \\
400 & 53.562 -i0.162716 & 53.6435 -i0.162373 & 56.812 -i0.162917 & 56.9245 -i0.161795  \\
 \hline
\end{tabular}%
\caption{QNMs calculated using the 6th-order WKB approximation method and the corresponding shadow radius.}
\label{tab5}
\end{table}

\section{Conclusion}

\label{sec6}
In this work, we construct a novel class of spherically symmetric black hole solutions in 4D noncommutative EGB gravity with a global monopole. The effects of the noncommutativity of spacetime are incorporated by modeling
the source of matter through a Lorentzian-smeared mass distribution. We then derive the metric function and analyze its horizon structure, demonstrating three distinct configurations: two horizons ($M>M_{c}$), a degenerate horizon ($M=M_{c}$), or a horizonless spacetime ($M<M_{c}$), where critical mass $M_{c}$ depends on the noncommutative parameter $\Theta $, symmetry-breaking scale $\eta $, and GB coupling constant $\alpha $. 

The thermodynamic analysis shows that the Hawking temperature, entropy, and heat capacity exhibit corrections from $\Theta $, $\eta $, and $\alpha $. The Hawking temperature and entropy acquire corrections from $\Theta $, $\eta$, and $\alpha $. The temperature exhibits a peak during evaporation, with its maximum shifting under variations of $\eta $ and $\alpha $. The heat capacity is negative for $r_{+}<r_{+}^{c}$, indicating an unstable
state, while the black hole is locally stable for $r_{+}>r_{+}^{c}$. These two states are separated by a second-order phase transition, where the heat capacity diverges at a critical radius $r_{+}^{c}$. The critical radius $
r_{+}^{c}$ is affected by variations in $\eta $ and $\alpha $. More precisely, it increases with the increase of $\alpha $ and decrease\ with $\eta $. 

Regarding the optical properties, we show that the black hole shadow radius decreases with increasing $\Theta $ or $\alpha $ but grows with $\eta $. These results reveal how noncommutative geometry and higher-curvature corrections modify light-bending phenomena, offering potential observational signatures for testing the model. 

Finally, the QNM frequencies of scalar perturbations, calculated using the sixth-order WKB method, exhibit stable damped oscillations. Both the real and imaginary parts of the frequencies vary monotonically with $\Theta $, $\alpha $, and $\eta $. The eikonal limit analysis confirms consistency between the shadow radius and QNM frequencies. 

This paper highlights the intricate interplay between noncommutative geometry, modified gravity, and topological defects, enriching our understanding of black hole thermodynamics and dynamics beyond general relativity. The dependence of the black hole shadow on the parameters $\Theta $, $\alpha $, and $\eta$ may help guide observational efforts to constrain these values. Future work can extend this analysis by exploring fermionic perturbations or interactions with dark matter, which can provide additional insights into the astrophysical relevance of these objects.
\section*{Conflict of Interests}

The authors declare no such conflict of interest.

\section*{Data Availability Statement}

No data were generated or analyzed in this study.

\end{document}